\documentclass[aps,prl,nofootinbib,twocolumn,showpacs,superscriptaddress,letterpaper,amsmath,amssymb]{revtex4}
\usepackage{graphicx}  % needed for figures
\usepackage{dcolumn}   % needed for some tables
\usepackage{bm}        % for math
\usepackage{amssymb}   % for math
\usepackage{feynmf}%%%{feynmp}
\usepackage{slashed}
\usepackage{multirow}
\usepackage{color}
\usepackage{array}

\newcolumntype{L}[1]{>{\raggedright\let\newline\\\arraybackslash\hspace{0pt}}m{#1}}
\newcolumntype{C}[1]{>{\centering\let\newline\\\arraybackslash\hspace{0pt}}m{#1}}
\newcolumntype{R}[1]{>{\raggedleft\let\newline\\\arraybackslash\hspace{0pt}}m{#1}}

\def\gsim{\lower0.5ex\hbox{$\:\buildrel >\over\sim\:$}}
\def\lsim{\lower0.5ex\hbox{$\:\buildrel <\over\sim\:$}}

\newcommand{\be}{\begin{equation}}
\newcommand{\ee}{\end{equation}}
\newcommand{\bea}{\begin{eqnarray}}
\newcommand{\eea}{\end{eqnarray}}

\newcommand{\nbox}{{\,\lower0.9pt\vbox{\hrule \hbox{\vrule height 0.2 cm
\hskip 0.2 cm \vrule height 0.2 cm}\hrule}\,}}

\def\sub#1{_{\lower.25ex\hbox{$\scriptstyle#1$}}}

\newskip\zatskip \zatskip=0pt plus0pt minus0pt
\def\matth{\mathsurround=0pt}
\def\lsim{\mathrel{\mathpalette\atversim<}}
\def\gsim{\mathrel{\mathpalette\atversim>}}
\def\sigv{\ifmmode \langle\sigma v\rangle\else $\langle\sigma v\rangle$\fi}
\newskip\zatskip \zatskip=0pt plus0pt minus0pt
\def\matth{\mathsurround=0pt}
\def\lsim{\mathrel{\mathpalette\atversim<}}
\def\gsim{\mathrel{\mathpalette\atversim>}}
\def\atversim#1#2{\lower0.7ex\vbox{\baselineskip\zatskip\lineskip\zatskip
  \lineskiplimit
  0pt\ialign{$\matth#1\hfil##\hfil$\crcr#2\crcr\sim\crcr}}}

\begin{document}

\thispagestyle{empty}
\vspace*{-3.5cm}

\vspace{0.5in}

%\begin{flushright}
%\today\\
%\end{flushright}
%\vspace{0.5in}
\title{Jet Substructure Classification in High-Energy Physics with Deep Neural Networks}

\begin{center}
\begin{abstract}
  At the extreme energies of the Large Hadron Collider,  massive particles can be produced at such high velocities that their hadronic decays are collimated and the resulting jets overlap.  Deducing whether the substructure of an observed jet is due to a low-mass single particle or due to multiple decay objects of a massive particle is an important problem in the analysis of collider data.  Traditional approaches have relied on expert features designed to detect energy deposition patterns in the calorimeter, but the complexity of the data make this task an excellent candidate for the application of machine learning tools. The data collected by the detector can be treated as a two-dimensional image, lending itself to the natural application of image classification techniques. In this work, we apply deep neural networks with a mixture of locally-connected and fully-connected nodes. Our experiments demonstrate that without the aid of expert features, such networks match or modestly outperform the current state-of-the-art approach for discriminating between jets from single hadronic particles and overlapping jets from pairs of collimated hadronic particles, and that such performance gains persist in the presence of pileup interactions.
\end{abstract}
\end{center}

\author{Pierre Baldi}
\affiliation{Department of Computer Science, University of  California, Irvine, CA 92697}
\author{Kevin Bauer}
\affiliation{Department of Physics and Astronomy, University of  California, Irvine, CA 92697}
\author{Clara Eng}
\affiliation{Department of Chemical Engineering, University of California Berkeley, Berkeley CA 94270}
\author{Peter Sadowski}
\affiliation{Department of Computer Science, University of  California, Irvine, CA 92697}
\author{Daniel Whiteson}
\affiliation{Department of Physics and Astronomy, University of  California, Irvine, CA 92697}

\date{\today}

\pacs{}
\maketitle

% introduction
%\linenumbers

\section{Introduction}

Collisions at the LHC occur at such high energies that even massive particles are produced at large enough velocities that their decay products  become collimated. In the case of a hadronic decay of a boosted $W$ boson ($W\rightarrow qq'$), the two jets produced from these two quarks then overlap in the detector, creating a single merged jet. The substructure of the jet's energy deposition can distinguish between jets which are due to a single hadronic particle or due to the decay of a massive object into multiple hadronic particles; this classification is known as jet ``tagging" and is critical for understanding the nature of the particles produced in the collision~\cite{Butterworth:2008iy}. 

This classification task has been the topic of intense research activity~\cite{Adams:2015hiv,Abdesselam:2010pt,Altheimer:2012mn,Altheimer:2013yza}. The difficult nature of the problem has lead physicists to reduce the dimensionality of the problem by designing  expert features~\cite{Plehn:2010st,Kaplan:2008ie,Larkoski:2014wba,Thaler:2010tr,Larkoski:2013eya,trim,prun,modified_mass,Dasgupta:2013via,Dasgupta:2015yua} which incorporate their domain knowledge. In the current state of the art applications, jets are either classified based on one of these features alone or by combining multiple designed features with shallow machine learning classifiers such as boosted decision trees (BDTs). It is possible, however, that these designed expert features do not capture all of the available information~\cite{baldi_searching_2014, baldi_enhanced_2015,sadowski_deep_2014}, as the data are very high-dimensional and despite extensive theoretical progress in the microphysics of jet formation~\cite{Soper:2011cr,Soper:2012pb,Stewart:2014nna} and the existence of effective simulation tools~\cite{pythia,Bahr:2008pv}, there exists no complete analytical model for classification directly from theoretical principles, though see Ref.~\cite{Larkoski:2015kga}.  Therefore, approaches that use the higher-dimensional but lower-level detector information to learn this classification function may outperform those which rely on fewer high-level expert-designed features.

Measurements of the emanating particles can be projected onto a cylindrical detector and then unwrapped and considered as two-dimensional images, enabling the natural application of computer vision techniques. Recent work demonstrates encouraging results with shallow classification models trained on jet images \cite{cogan_jet_images_2015,almeida_playing_2015,boosted_w}.  Deep networks have shown additional promise in particle-level studies~\cite{deOliveira:2015xxd}. However, deep learning has not yet been applied to more realistic scenarios which include simulation of the detector response and resolution, and most importantly, the effect of unrelated simultaneous $pp$ interactions, known as {\it pileup} which contributes significant energy depositions unrelated to the particles of interest.

In this paper, we perform jet classification on images built from simulated detector response using deep neural network models with a combination of locally-connected and fully-connected layers.  Our results demonstrate that deep networks can distinguish between detector clusters due to single or multiple jets without using domain knowledge, matching or exceeding the performance of shallow classifiers used to combine many expert features.

\section{Theory}

A typical application of jet classifiers is to discriminate single jets produced in quark or gluon fragmentation from two overlapping jets produced when a high-velocity $W$ boson decays to a collimated pair of quarks. The goal is then to learn the classification function, or equivalently, the likelihood ratio:

\[ \frac{P_{W\rightarrow qq}(\textrm{jet})}{P_{q/g}(\textrm{jet})} \]

In practice, there are two significant obstacles to calculating and applying this ratio.  

First, while theoretical understanding of the processes involved has made significant progress, a formulation of this likelihood ratio from fundamental QCD principles is not yet available.  However, there do exist effective models which have been successfully incorporated into widely used tools capable of generating simulated samples. Such samples can then be used to deduce the likelihood ratio, but the task is very difficult due to its high-dimensionality. Expert features with solid theoretical grounding exist to reduce the dimensionality of this problem, but it is unlikely that they capture all of the information, as the theoretical understanding is not complete and the concepts which motivate them do not include the detector effects or the impact of pileup interactions.  The goal of this paper is to attempt to capture as much of the information as possible and learn the classification function from  simulated samples which include these effects, without making the simplifying theoretical assumptions necessary to construct expert features.

Second, the effective models used in simulation tools do not provide a perfectly accurate description of observed collider data.  A classification function learned from simulated samples is limited by the validity of those samples. While deep networks may provide a powerful method of deducing the classification function, expert features which encapsulate theoretical understanding of the process of jet formation are valuable in assessing the success and failure of these models.  In this paper, we use expert features as a benchmark to measure the performance of learning tools which access only the higher-dimensional lower-level data. We expect that deep networks may provide additional classification power in concert with the insight offered by expert features, and perhaps motivate the development of modifications to such features rather than blindly replacing them.

\section{Data}

Training samples for both classes were produced using realistic simulation tools widely used in particle physics.

 Samples of boosted $W\rightarrow qq'$ were generated with a center of mass energy $\sqrt{s}=14$~TeV using the diboson production and decay process $pp\rightarrow W^{+}W^{-}\rightarrow qqqq$ leading to two pairs of quarks; each pair of quarks are collimated and lead to a single jet.   Samples of jets originating from single quarks and gluons were generated using the  $pp\rightarrow qq,qg,gg$ process. In both cases, jets are generated in the range of $p_{\textrm{T}} \in [300,400]$ GeV.

Collisions and immediate decays were simulated  with {\sc{madgraph5}}~\cite{madgraph} v2.2.3, showering and hadronization simulated with {\sc pythia}~\cite{pythia} v6.426 , and response of the detectors simulated with {\sc delphes}~\cite{delphes} v3.2.0. The jet images are characterized by the energies deposited at different points on the approximately cylindrical calorimeter surface.

The classification of jets as due to $W\rightarrow qq'$ or single quarks and gluons is sensitive to the presence of additional in-time $pp$ interactions, referred to as {\it pile-up} events.  We overlay such interactions in the simulation chain, with an average number of interactions per event of $\left<\mu\right>=50$, as an estimate of future ATLAS Run 2 data with the LHC delivering collisions at a 25ns bunch crossing interval. The impact of pile-up events on jet reconstruction can be mitigated using several techniques.
After reconstructing  jets with the anti-$k_{\textrm{T}}$~\cite{Cacciari:2008gp} clustering algorithm using distance parameter $R=1.2$, we apply a jet-trimming algorithm~\cite{Krohn:2009th} which is designed to remove pileup while preserving the two-pronged jet substructure characteristic of boson decay. Jet trimming re-clusters the jet constituents using the $k_{\textrm{T}}$~\cite{Ellis:1993tq}   algorithm into subjets of radius 0.2 and discards subjets with $p_{\textrm{T}}$ less than 3\% of the original jet. Then the final trimmed jet is built using the remaining subjets. Trimmed jets with 300 GeV$<p_{\textrm{T}}<$400 GeV are selected, in order to ensure the minimum $W$ boson velocity needed for collimated decays. In principle, the machine learning algorithms may be able to classify jets without such filtering; we leave this for future studies.

\begin{figure}
\begin{center}
\includegraphics[width=0.45\linewidth]{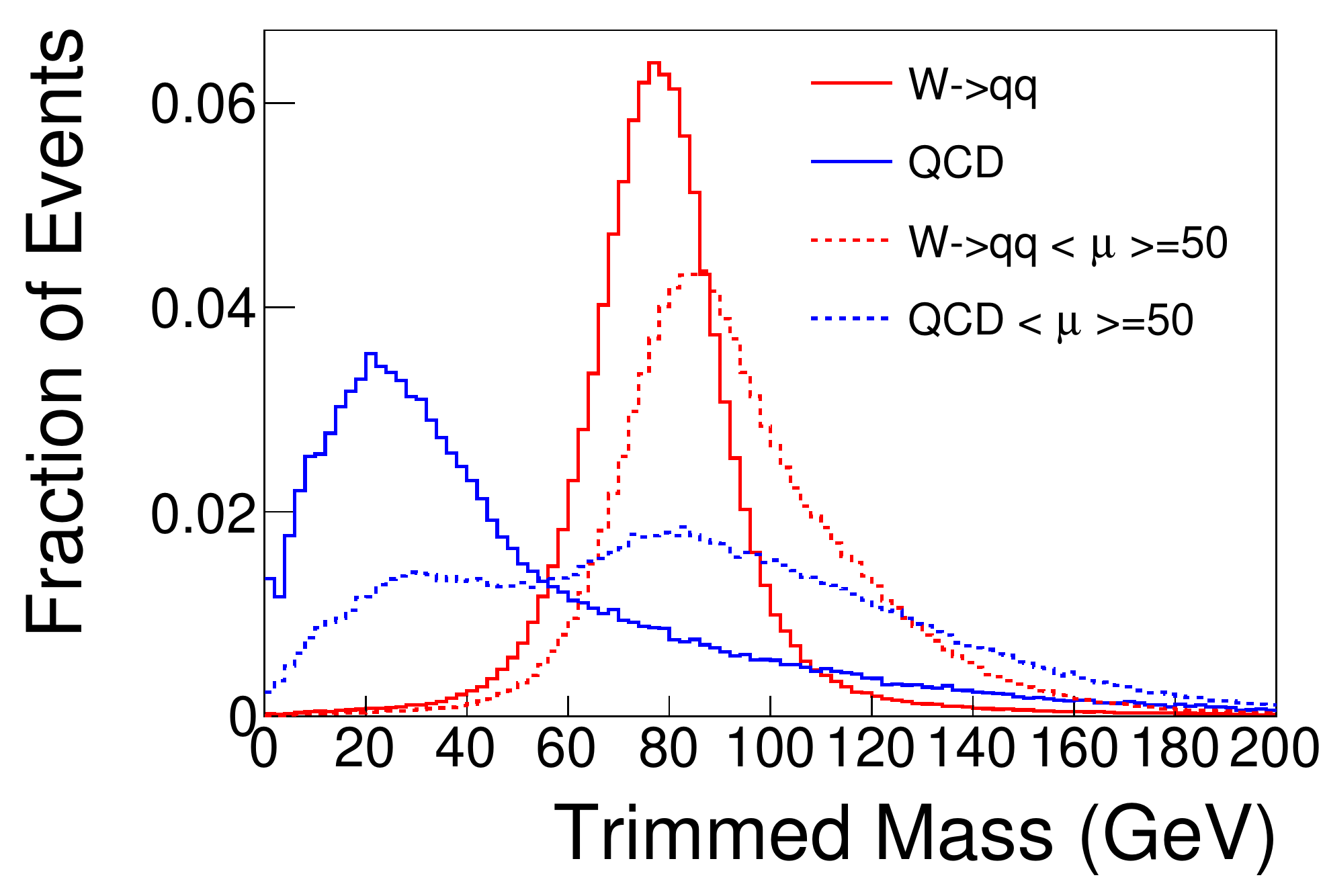}
\includegraphics[width=0.45\linewidth]{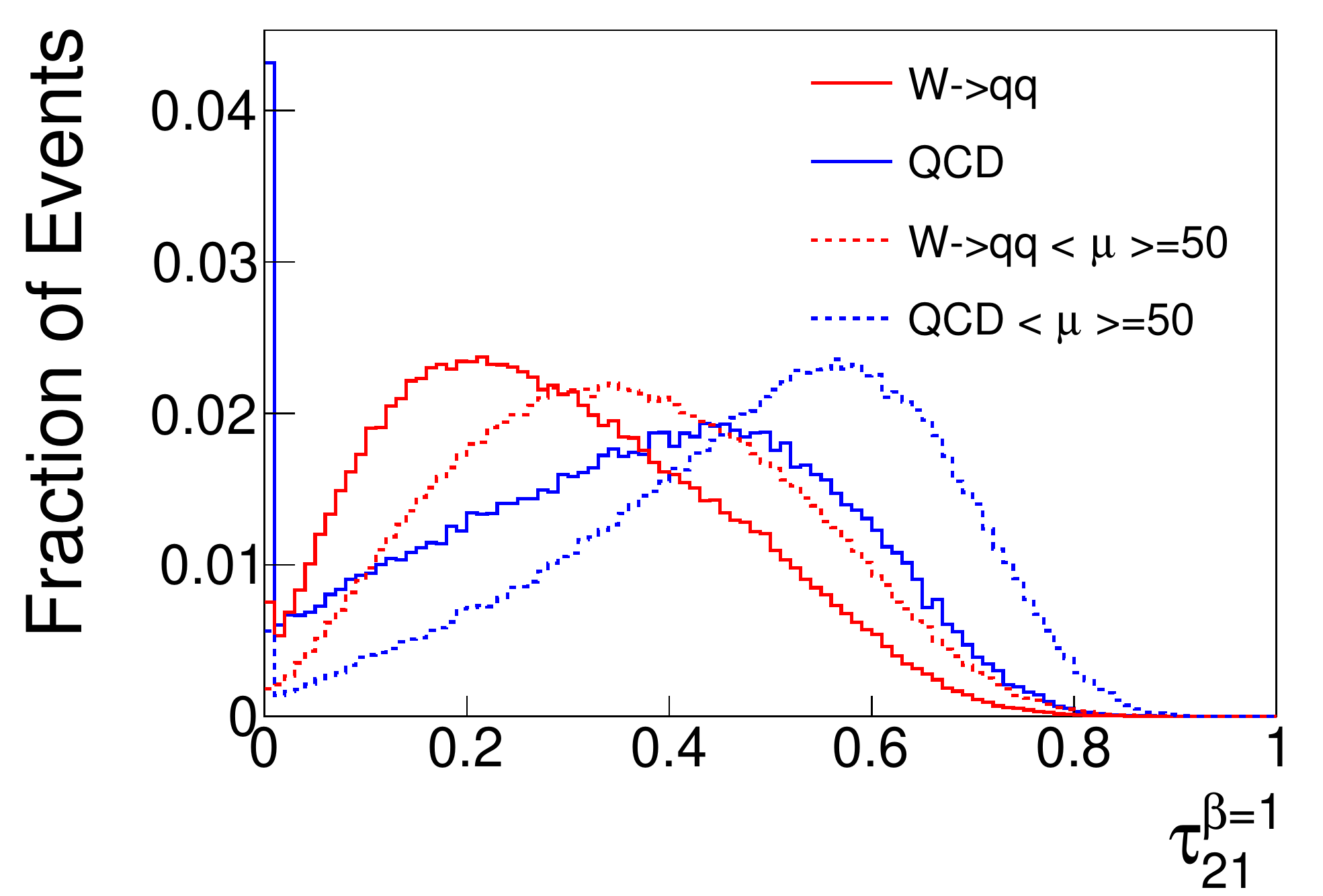}
\includegraphics[width=0.45\linewidth]{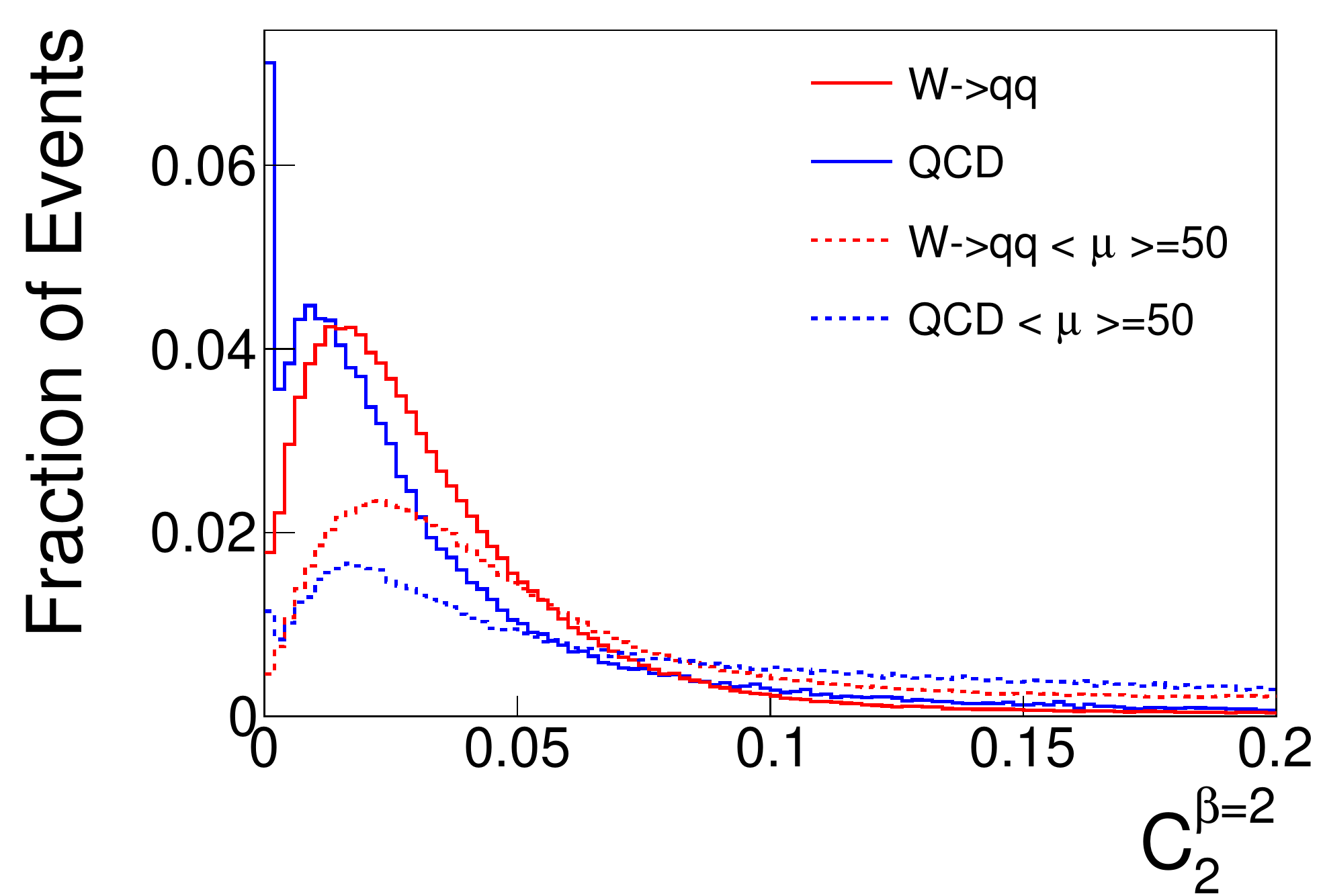}
\includegraphics[width=0.45\linewidth]{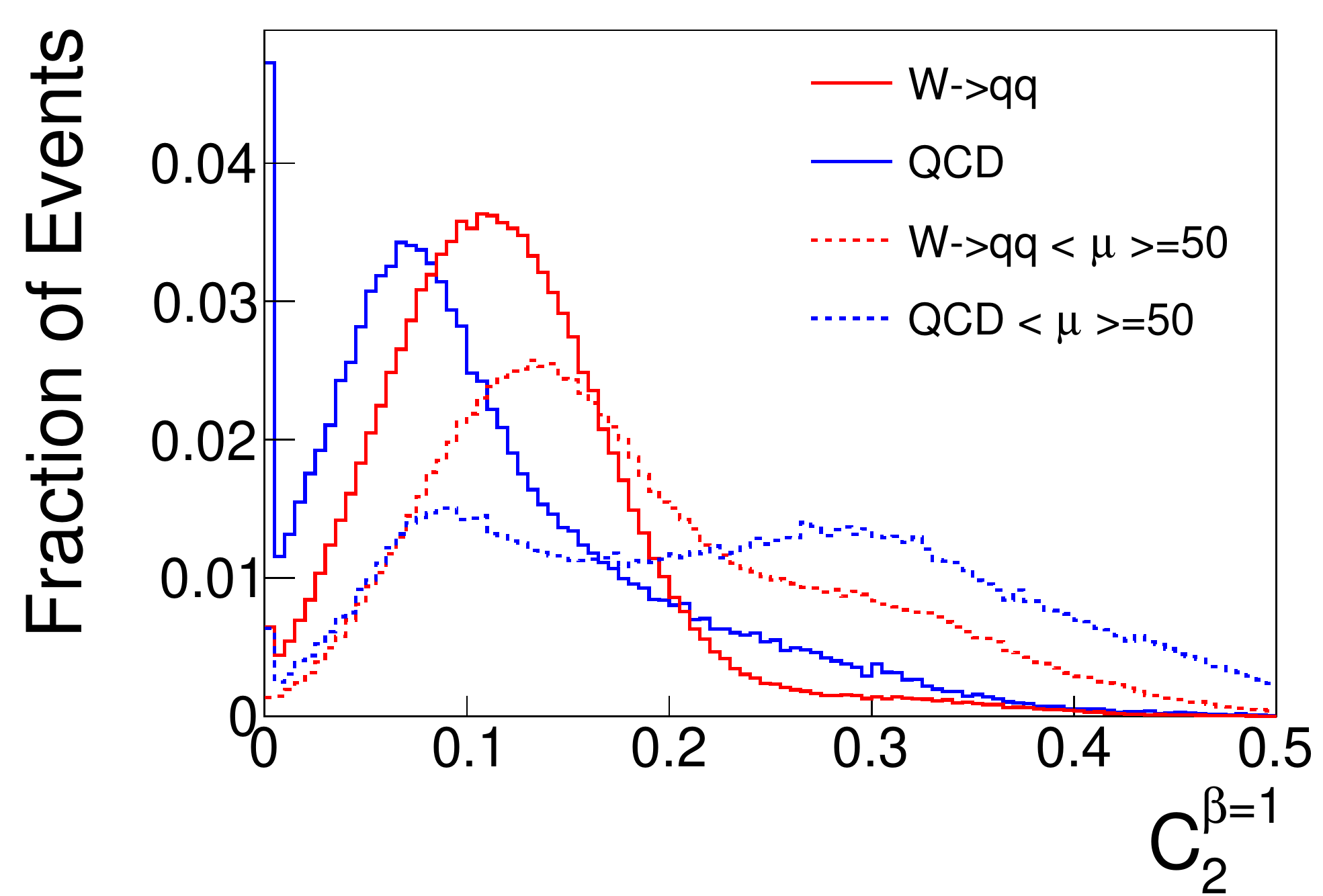}
\includegraphics[width=0.45\linewidth]{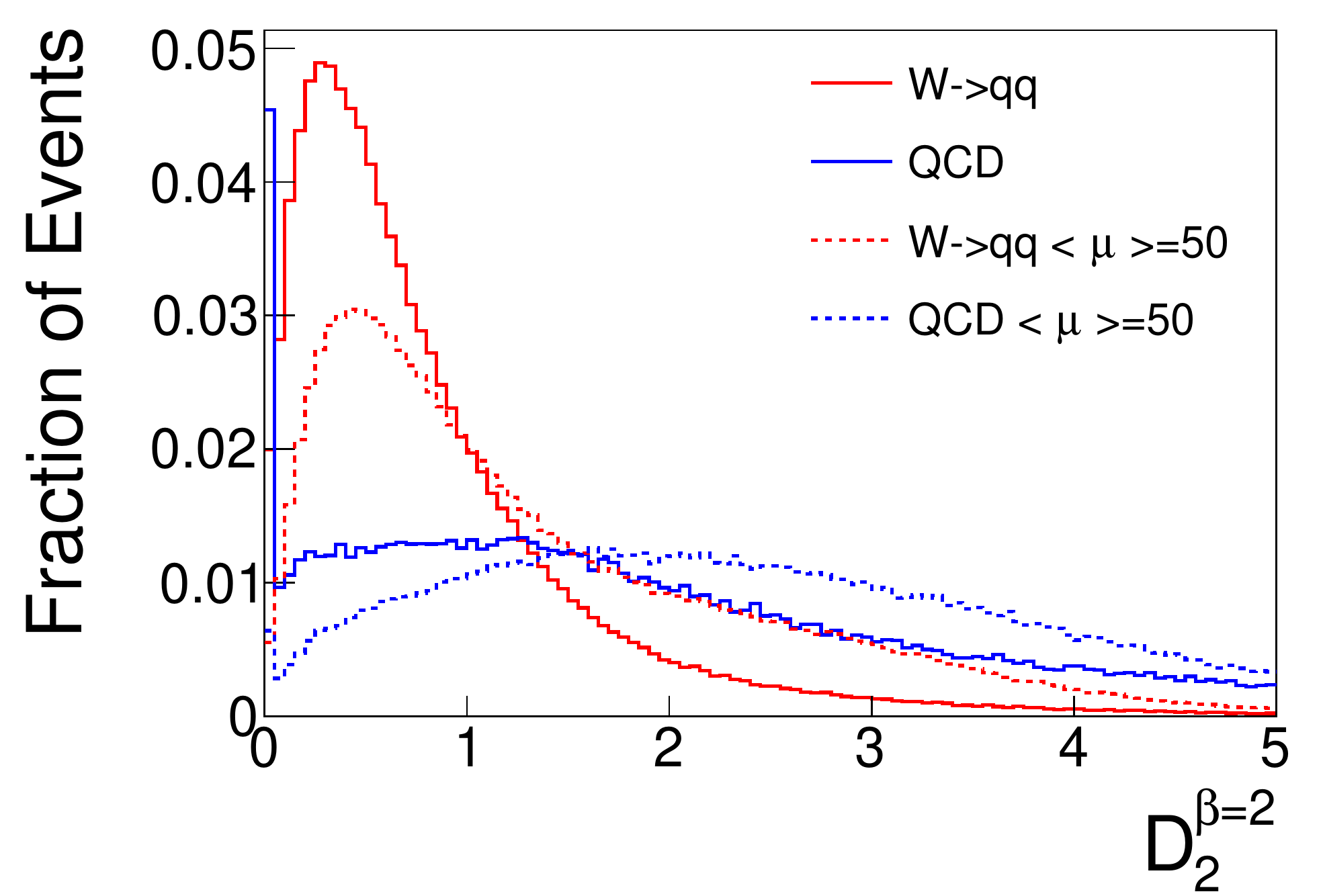}
\includegraphics[width=0.45\linewidth]{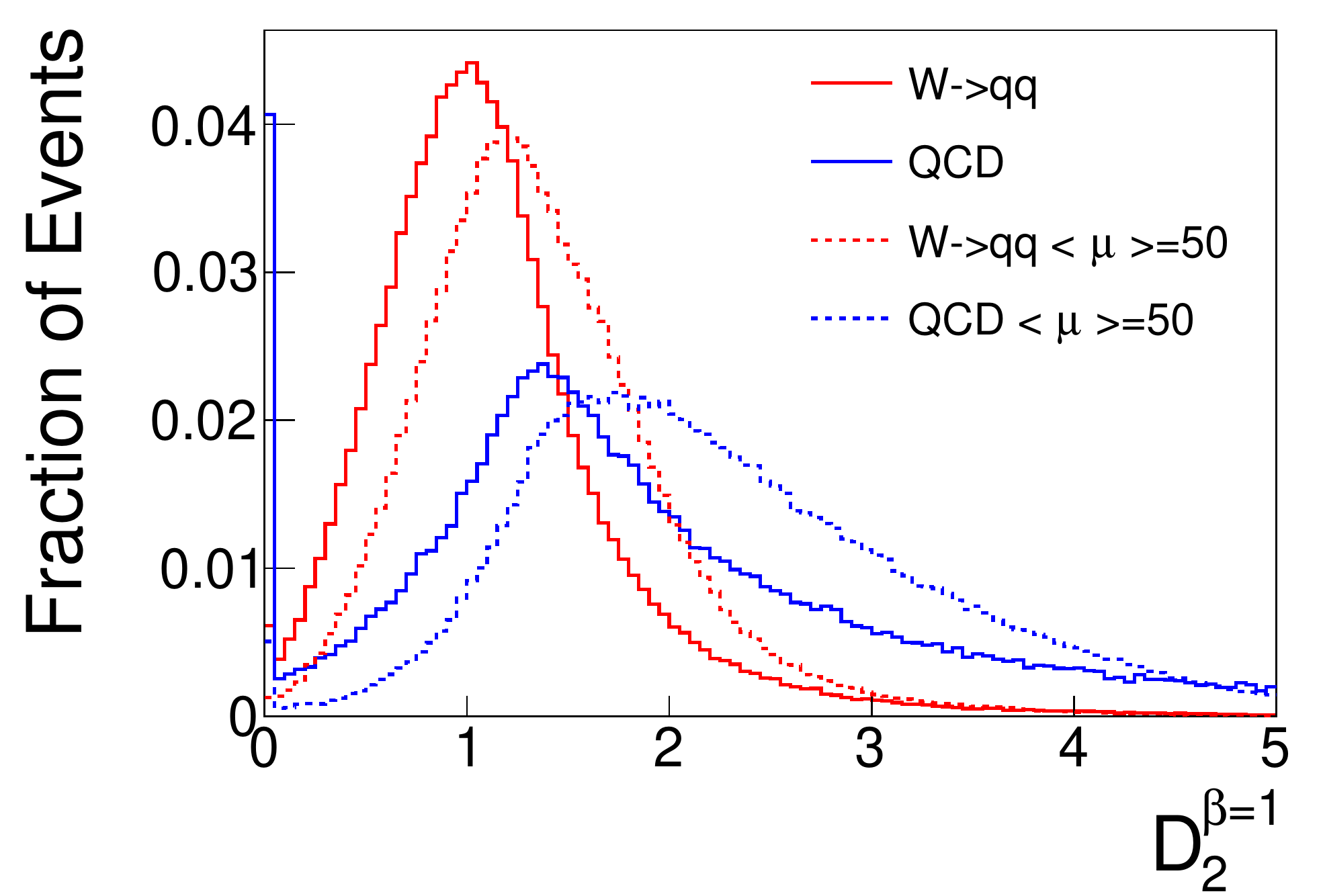}
\end{center}
\caption{Distributions in simulated samples of  high-level jet substructure variables widely used to discriminate between jets due to collimated decays of massive objects ($W\rightarrow qq$) and jets due to individual quarks or gluons (QCD).  Two cases are shown: with and without the presence of additional in-time $pp$ interactions, included at the level of an average of 50 such interactions per collision.}
\label{fig:var}
\end{figure}

To compare our approach to the current state-of-the-art, we calculate six high-level jet variables commonly used in the literature; calculations are performed using FastJet~\cite{Cacciari:2011ma} v3.1.2. First, the invariant mass of the trimmed jet is calculated. Then, the trimmed jet's constituents are used to calculate the other substructure variables, $N$-subjettiness~\cite{Thaler:2010tr,Thaler:2011gf} $\tau^{\beta=1}_{21}$, and the energy correlation functions~\cite{Larkoski:2013eya,Larkoski:2014gra} $C^{\beta=1}_{2}$, $C^{\beta=2}_{2}$, $D^{\beta=1}_{2}$, and $D^{\beta=2}_{2}$. A comprehensive summary of these six jet substructure variables can be found in Ref.~\cite{Adams:2015hiv}.  Figures~\ref{fig:var} shows the distribution of the variables for the two classes of jets, both  with and without pileup conditions.

In this paper, we investigate the power of classification of the jets directly from the lower-level but higher-dimensional calorimeter data, without the dimensional reduction provided by the variables above.  The strategy follows that of well-established image classification tools by treating the distribution of energy in the calorimeter as an image. The images were preprocessed as in previous work by centering and rotating into a canonical orientation. The origin of the coordinate axis was set at the center of energy of each jet, then the image was rotated so that the principle axis $\theta$ is in the same direction for each jet, where $\theta$ is defined as

\begin{align}
\tan(\theta) &= \sum_i \frac{\phi_i \times E_i}{R_i} \bigg/ \sum_i \frac{\eta_i \times E_i}{R_i} \\
R_i &= \sqrt{\eta_i^2 + \phi_i^2}.
\end{align}

Images are then reflected so that the maximum energy value is always in the top half of the image. 

\begin{figure}
\begin{center}
\includegraphics[width=0.45\linewidth]{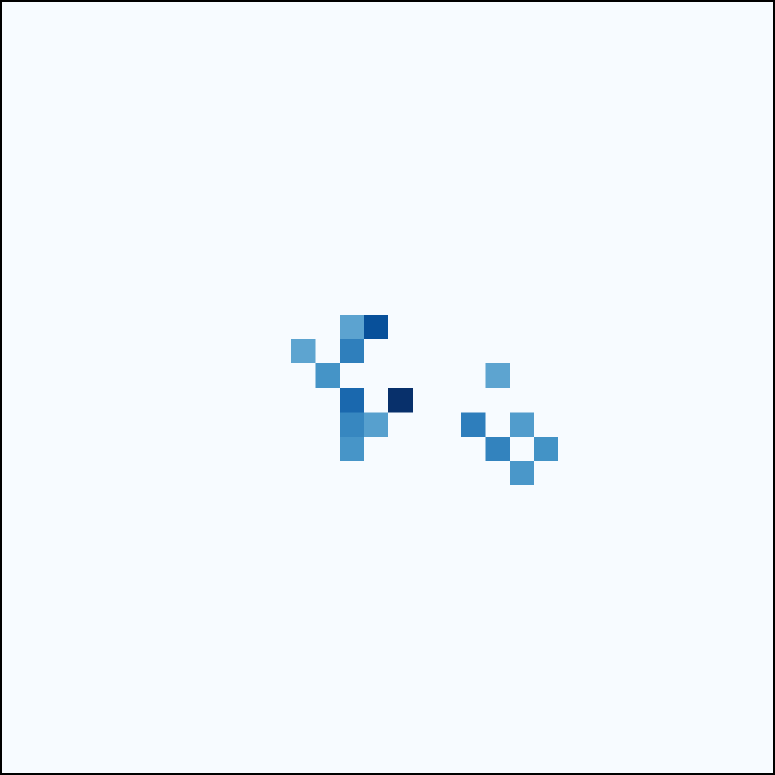}
%\caption*{Example image from class 1 (a single jet), after preprocessing.}
\includegraphics[width=0.45\linewidth]{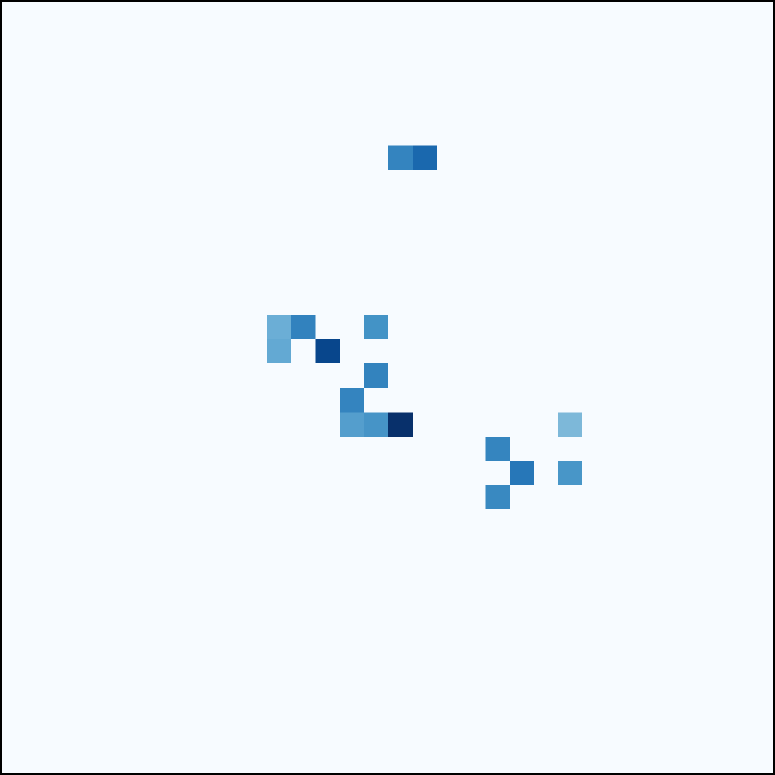}
\end{center}
\caption{Typical jet images from class 1 (single QCD jet from $q$ or $g$) on the left, and class 2 (two overlapping jets from $W\rightarrow qq'$) on the right, after preprocessing as described in the text.}
\label{fig:jet_examples}
\end{figure}

\begin{figure}
\begin{center}
\includegraphics[width=0.45\linewidth]{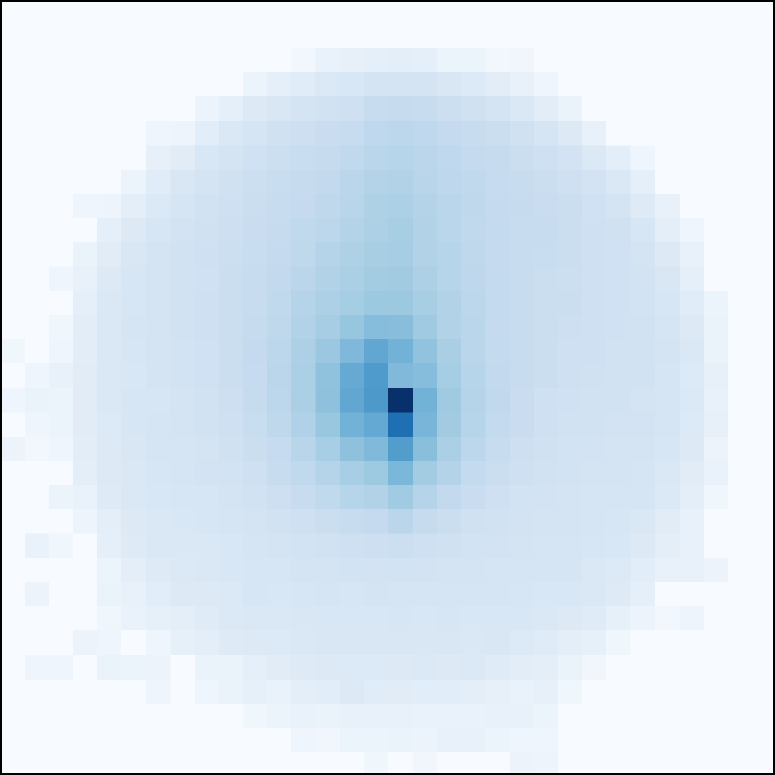}
\includegraphics[width=0.45\linewidth]{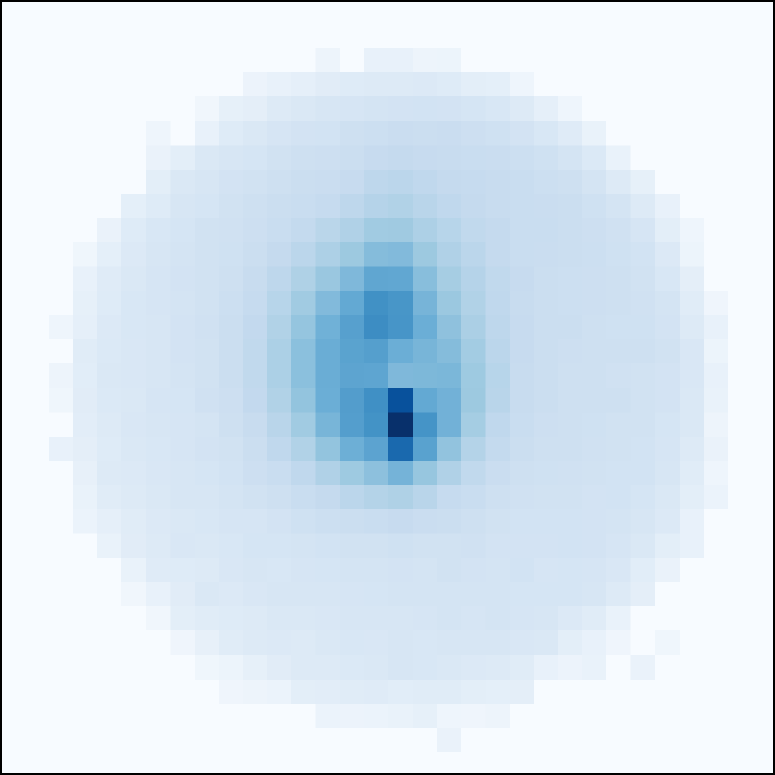}
\end{center}
\caption{Average of 100,000 jet images from class 1 (single QCD jet from $q$ or $g$) on the left, and class 2 (two overlapping jets from $W\rightarrow qq'$) on the right, after preprocessing.}
\label{fig:jet_means}
\end{figure}

The jet energy deposits were centered and cropped to within a $3.0\times3.0$ radian window, then binned into pixels to form a $32\times32$ image, approximating the resolution of the calorimeter cells. When two calorimeter cells were detected within the same pixel, their energies were summed. Example individual jet images from each class are shown in Figure \ref{fig:jet_examples}, and averages over many jets are shown in Figure~\ref{fig:jet_means}. 

% We have to include information about the high level features used to train the boosted decision tree. Note we only included cases in which all the classifiers were able to produce an output.

%%%%%%%%%%%%%%%%%%%%%%%%%%%%%%%%%%%%%%%%%%%%%%%%%%%%%%%%%%%%%%%%%%%%%%%%%%%%%%%%%%%%
\section{Training}

Deep neural networks were trained on the jet images and compared to the standard approach of BDTs trained on expert-designed variables that capture domain knowledge~\cite{Adams:2015hiv}. All classifiers were trained on a balanced training data set of 10 million examples, with 500 thousand of these used as a validation set. The best hyperparameters for each method were selected using the Spearmint Bayesian optimization algorithm~\cite{snoek_practical_2012} to optimize over the supports specified in Tables \ref{tab:hp} and \ref{tab:bdt_support}. The best models were then tested on a separate test set of 5 million examples. 

Neural networks consisted of hidden layers of $tanh$ units and a logistic output unit with cross-entropy loss. Weight updates were made using the ADAM optimizer~\cite{kingma_adam:_2014} ($\beta_1=0.9, \beta_2=0.999, \epsilon=1e-08$) with mini-batches of size 100. Weights were initialized from a normal distribution with the standard deviation suggested by Ref.~\cite{he_delving_2015}. The learning rate was initialized to $0.0001$ and decreased by a factor of $0.9$ every epoch. Training was stopped when the validation error failed to improve or after a maximum of 50 epochs. All computations were performed using Keras~\cite{chollet_keras_2015} and Theano~\cite{bergstra_theano:_2010,bastien_theano:_2012} on NVidia Titan X processors. Convolutional networks were also explored, but as expected, the translational invariance provided by these architectures did not provide any performance boost.

We explore the use of locally-connected layers, where each neuron is only connected to a distinct 4-by-4 pixel region of the previous layer. This local connectivity constrains the network to learn spatially-localized features in the lower layers without assuming translational invariance, as in convolutional layers where the weights of the receptive fields are shared. Fully-connected layers were stacked on top of the locally-connected layers to aggregate information from different regions of the detector image. The network architecture --- the number of layers of each type, plus the width of the fully-connected layers --- was optimized using Spearmint. Out of the 25 network architectures explored on the no-pile-up task, the best had four locally-connected layers followed by four fully-connected layers of 425 units. This network has roughly 750,000 tunable parameters, while the best shallow network (one hidden layer of 1000 units) had over 1 million parameters. On the pile-up data, 19 different network architectures were tested; the best was again an 8-hidden-layer architecture, with 3 locally-connected layers, five fully-connected layers, and 500 hidden units in each layer.

BDTs were trained on the six high-level variables using Scikit-Learn~\cite{scikit-learn}. The maximum depth of each estimator, the minimum number of examples required to constitute an internal node (parameterized as a fraction of the training set), and the learning rate were separately optimized for the datasets with and without pileup using Spearmint (110 and 140 experiments, respectively). The number of estimators was fixed to 500; when evaluating the marginal improvement of performance with the addition of each estimator, we observed that in the best model, performance plateaued after inclusion of less than 100 estimators. This suggests that the number of estimators was not limiting. The minimum number of examples required to form a leaf node was fixed to be one fourth of that required to constitute an internal node. In both cases, the best BDT classifier had a maximum tree depth of 49, a minimum split requirement of 0.0021, and a learning rate of 0.07. The best BDT trained on the no-pileup data had approximately 700,000 tunable parameters, while the best BDT trained on the pileup data had approximately 750,000.

\section{Results}

%\begin{figure}
%\begin{center}
%\includegraphics[width=0.8\linewidth]{auc_vs_depth.pdf}
%\end{center}
%\caption{Effect of neural network depth on performance. For each number of total hidden layers (either locally-connected or fully-connected), we plot the test set performance of the best-performing classifier observed in the hyperparameter search. The best shallow network (one hidden layer) has 1 million parameters, more than the best deep network.}
%\label{fig:depth}
%\end{figure}

%The performance of the best architectures for each depth are plotted in Fig.~\ref{fig:depth}, showing t

Deep networks with locally-connected layers showed the best performance. For example, the best network with 5 hidden layers has two locally-connected layers followed by three fully-connected layers of 300 units each; this architecture performs better than a network of five fully-connected layers of 500 units each.

\begin{figure}
\begin{center}
\includegraphics[width=0.8\linewidth]{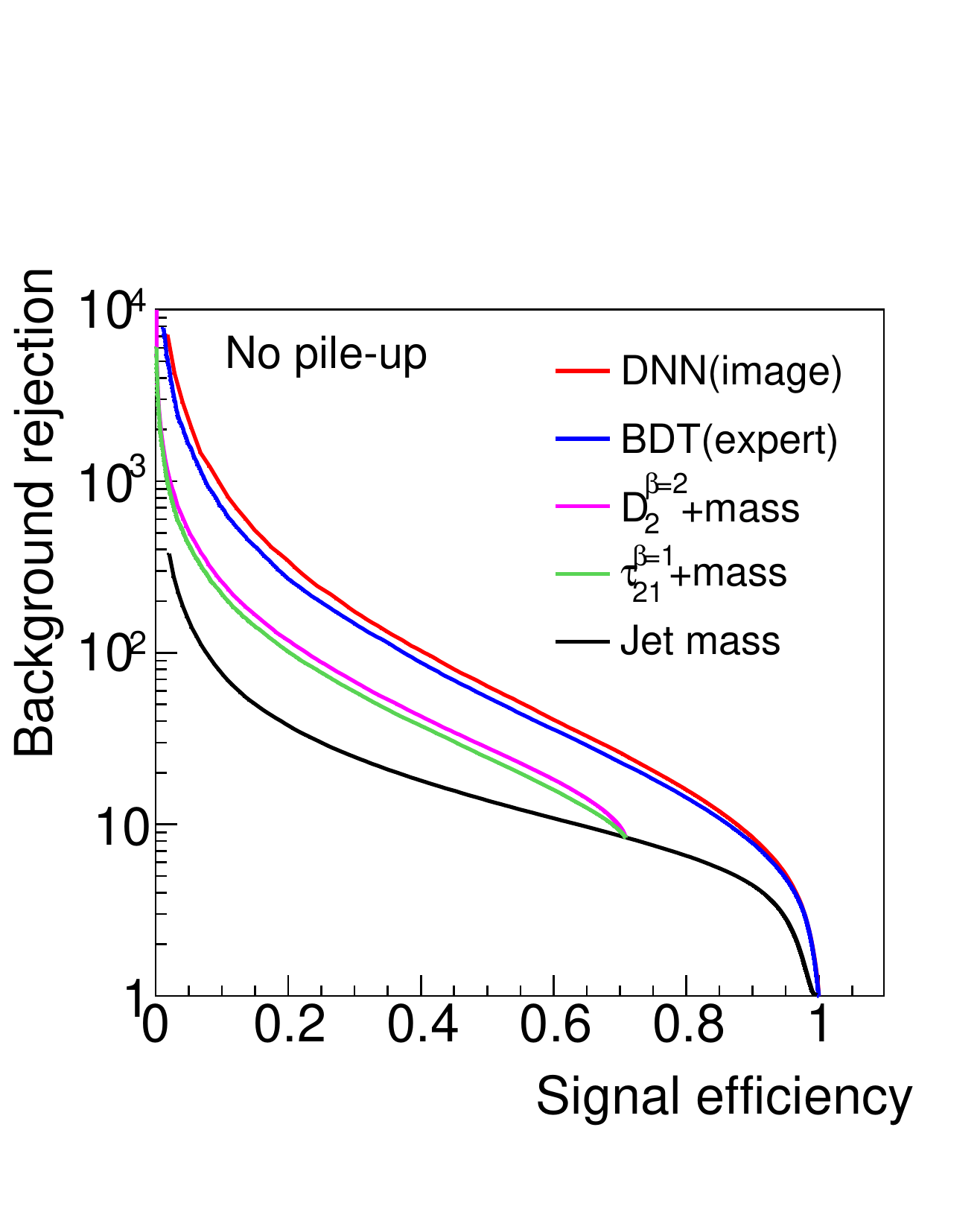}
\includegraphics[width=0.8\linewidth]{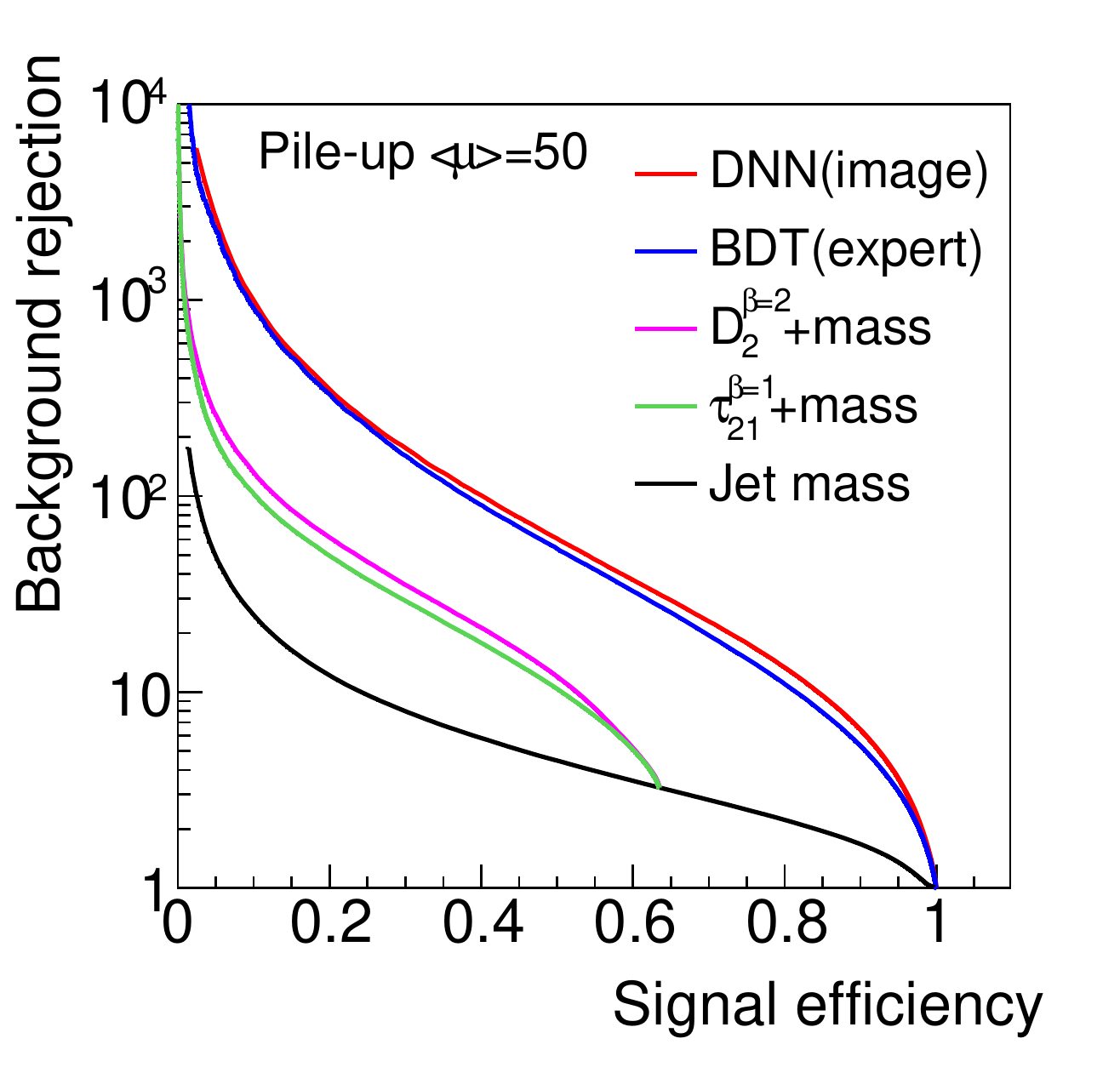}
\end{center}
\caption{Signal efficiency versus background rejection (inverse of efficiency) for deep networks trained on the images and boosted decision trees trained on the expert features, both with (bottom) and without pile-up (top). Typical choices of signal efficiency in real applications are in the 0.5-0.7 range. Also shown are the performance of jet mass individually as well as two expert variables in conjunction with a mass window.}
\label{fig:roc}
\end{figure}

Final results are shown in Table~\ref{tab:performance}. The metric used is the Area Under the Curve (AUC), calculated in signal efficiency versus background efficiency, where a larger AUC indicates better performance.  In Fig~\ref{fig:roc}, the signal efficiency is shown versus backround rejection, the inverse of background efficiency. In the case without pile-up, as studied in Ref.~\cite{deOliveira:2015xxd}, the deep network modestly outperforms the physics domain variables, demonstrating first that successful classification can be performed without expert-designed features and that there is some loss of information in the dimensional reduction such features provide.  See the discussion below, however, for comments on the continued importance of expert features.

 Our results also demonstrate for the first time that such performance holds up under the more difficult and realistic conditions of many pileup interactions; indeed, the gap between the deep network and the expert variables in this case is more pronounced. This is likely due to the fact that the physics-inspired variables rest on arguments motivated by idealized pictures.

\begin{table}
\centering
\caption{Hyperparameter support for Bayesian optimization of deep neural network architectures. For the no-pileup case, networks with a single hidden layer were allowed to have up to 1000 units per layer, in order to remove the possibility of the deep networks performing better simply because they had more tunable parameters.}
\begin{tabular}{lrrrr}
\hline\hline
&\multicolumn{2}{c}{Range} & \multicolumn{2}{c}{Optimum} \\
Hyperparameter & Min & Max & No pileup & Pileup \\
\hline
Hidden units per layer  & 100 & 500 & 425 & 500 \\
Fully-connected layers & 1 & 5 & 4 & 5 \\
Locally-connected layers & 0 & 5 & 4 & 3 \\
\hline\hline
\end{tabular}
\label{tab:hp}
\end{table}

\begin{table}
\centering
\caption{Hyperparameter support for BDTs trained on 6 high-level features, and the best combinations in 110 and 140 experiments, respectively, for the no-pileup and pileup tasks.  Minimum leaf percent was constrained to be one fourth of the minimum split percent in all cases.}
\begin{tabular}{lrrrr}
\hline\hline
&\multicolumn{2}{c}{Range} & \multicolumn{2}{c}{Optimum} \\
Hyperparameter & Min & Max & No pileup & Pileup  \\
\hline
Tree depth & 15 & 75 & 49 & 49 \\
Learning rate  & 0.01 & 1.00 & 0.07 & 0.07 \\
Minimum split percent & 0.0001 & 0.1000 & 0.0021 & 0.0021 \\
\hline\hline
\end{tabular}
\label{tab:bdt_support}
\end{table}

\begin{table}
\caption{ Performance results for BDT and deep networks. Shown for each method are both  the signal efficiency at background rejection of 10, as well as the Area Under the Curve (AUC), the integral of the background efficiency versus signal efficiency. For the neural networks, we report the mean and standard deviation of three networks trained with different random initializations.}
\begin{center}
\begin{tabular}{llll}
\hline\hline
 & \multicolumn{2}{c}{Performance} \\
Technique & Signal efficiency & AUC \\
& at  bg. rejection=10 & \\
\hline
& \multicolumn{2}{c}{{\it No pileup}}\\
BDT on derived features  		& $86.5\%$ & $95.0\%$ \\ 
Deep NN on images 	& $87.8\%${\tiny (0.04\%)} & $95.3\%${\tiny (0.02\%)} \\
\hline
& \multicolumn{2}{c}{{\it With pileup}}\\
BDT on derived features 					& $81.5\%$ & $93.2\%$ \\ 
Deep NN on images 		& $84.3\%${\tiny (0.02\%)} & $94.0\%${\tiny (0.01\%)} \\
\hline\hline
\end{tabular}
\end{center}
\label{tab:performance}
\end{table}

\section{Interpretation}

Current typical use in experimental analysis is the combination of the jet mass feature with $\tau_{21}$ or one of the energy correlation variables.  Our results show that even a straightforward BDT-combination of all six of the high-level variables provides a large boost in comparison.  In probing the power of deep learning, we then use as our benchmark this combination of the variables provided by the BDT.

The deep network has clearly managed to match or slightly exceed the performance of a combination of the state-of-the-art expert variables.  Physicists working on the underlying theoretical questions may naturally be curious as to whether the deep network has learned a novel strategy for classification which could inform their studies, or rediscovered and further optimized the existing  features. 

\begin{figure}
\begin{center}
\includegraphics[width=0.45\linewidth]{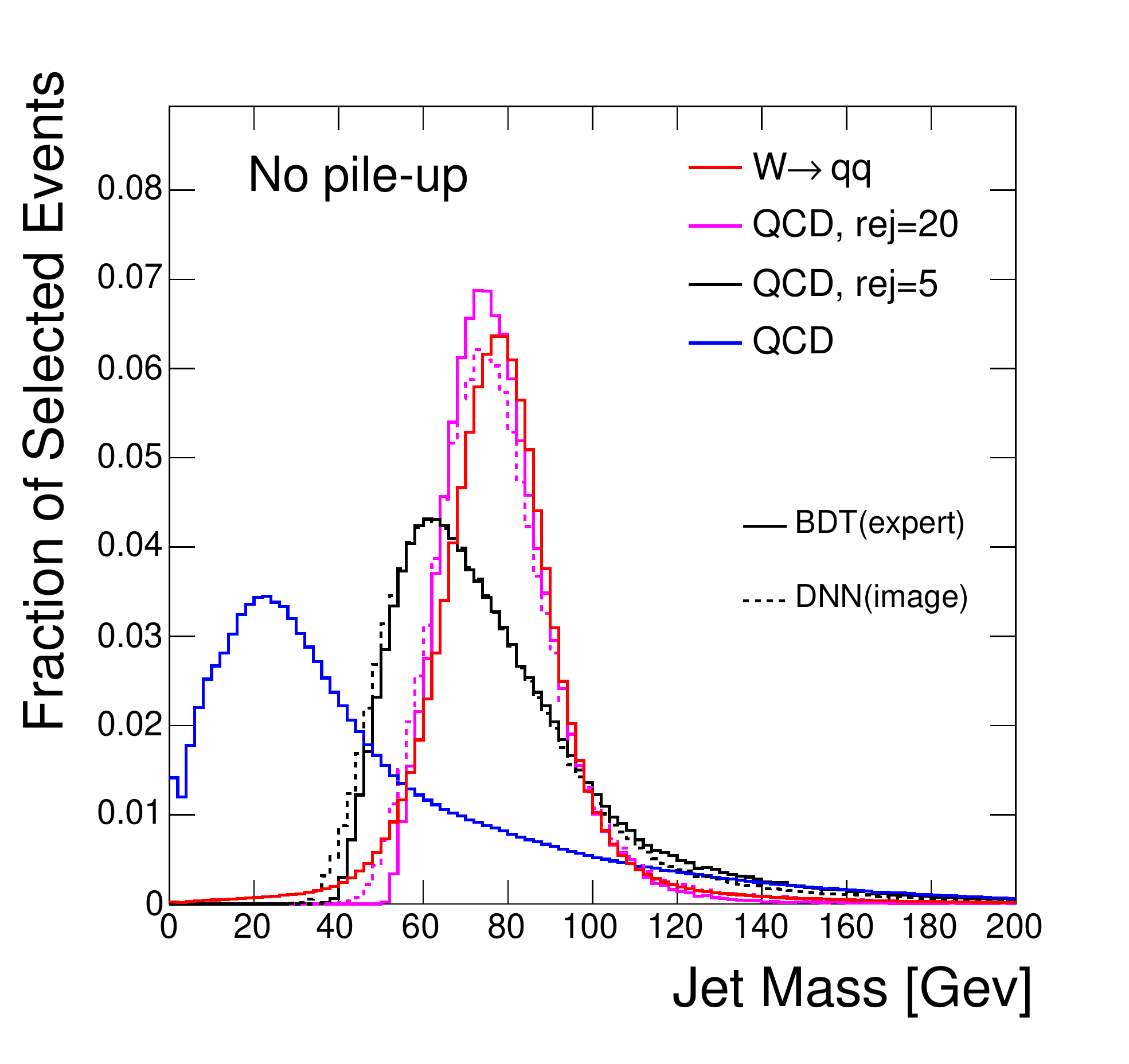}
\includegraphics[width=0.45\linewidth]{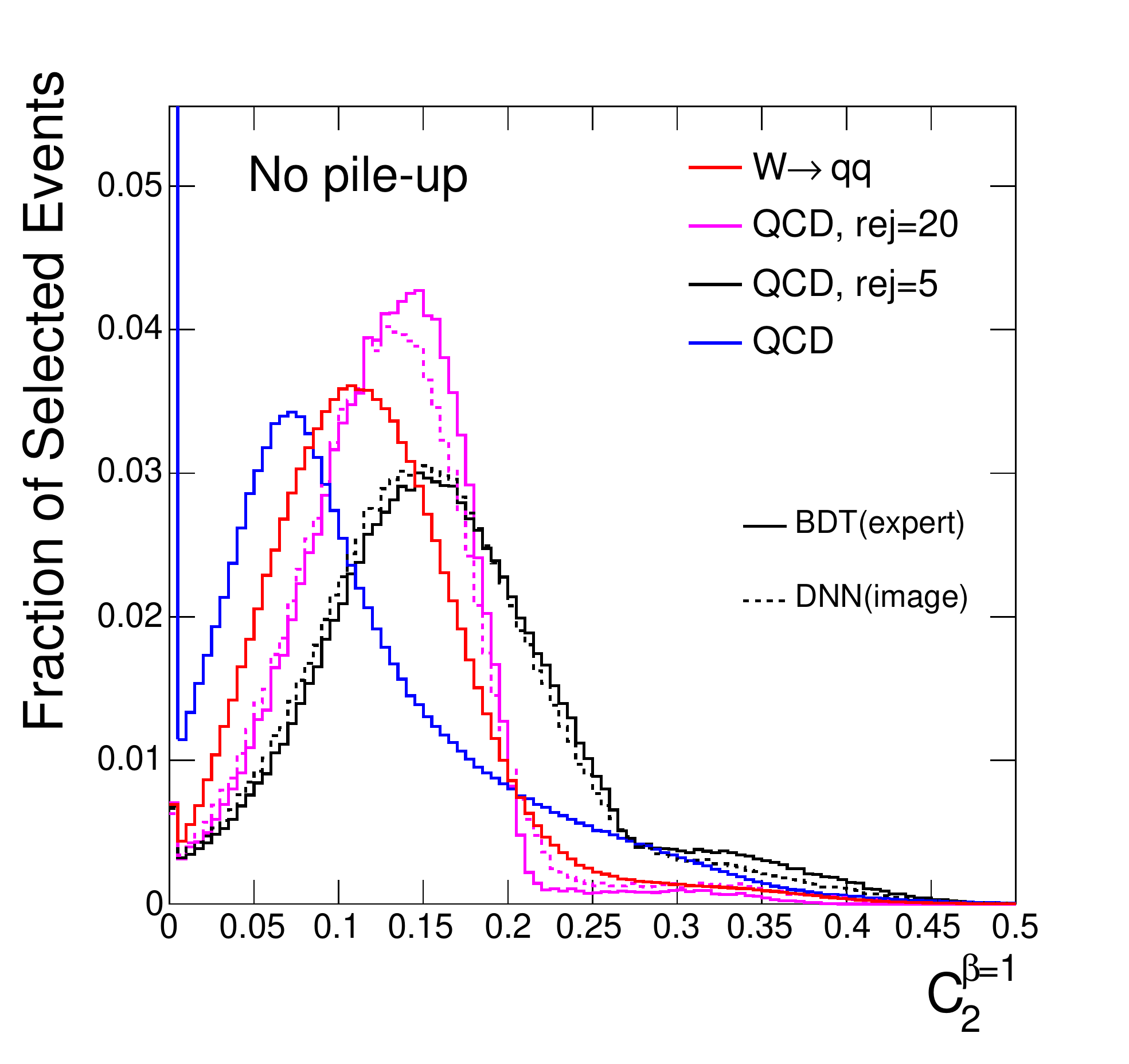}
\includegraphics[width=0.45\linewidth]{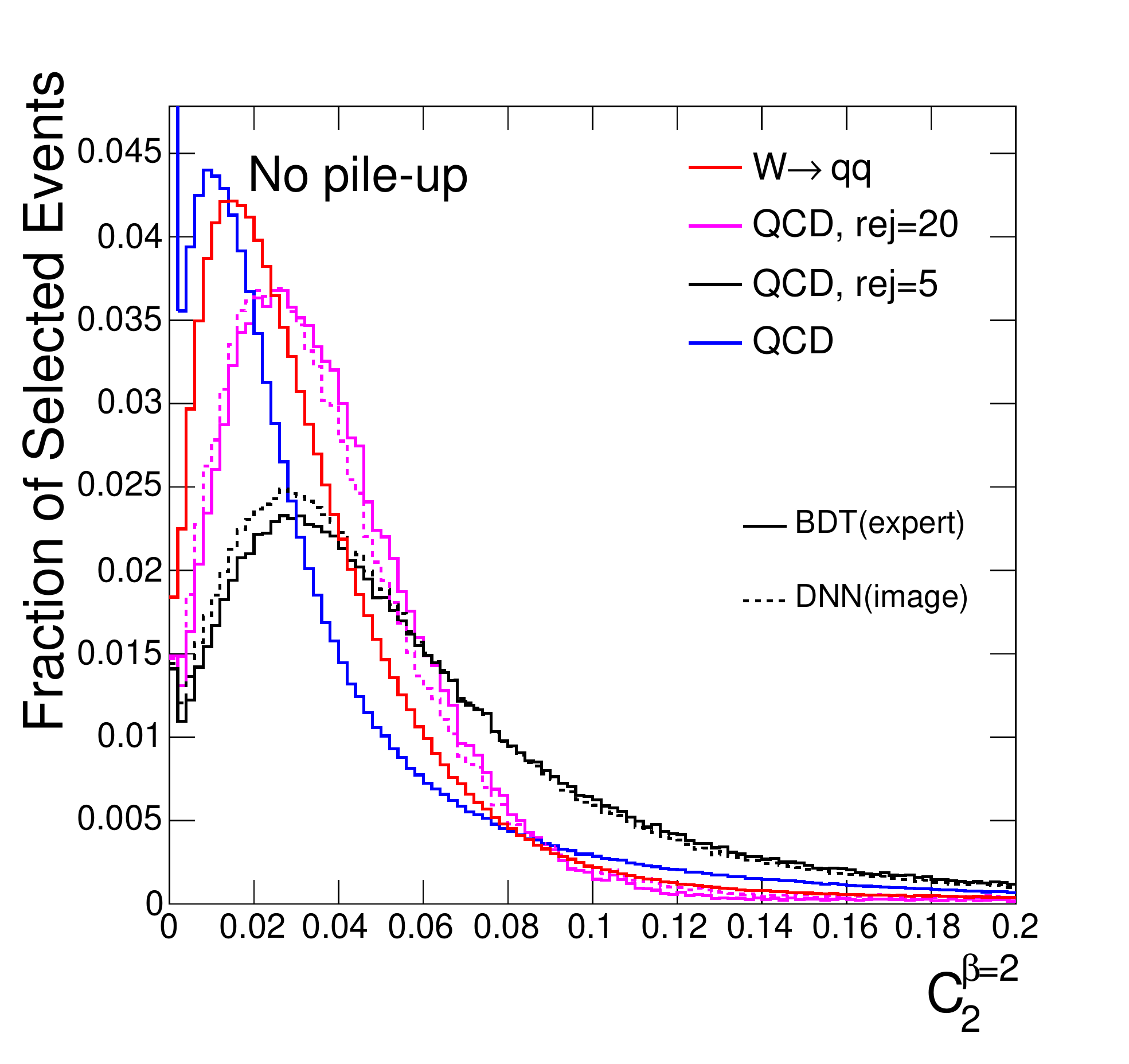}
\includegraphics[width=0.45\linewidth]{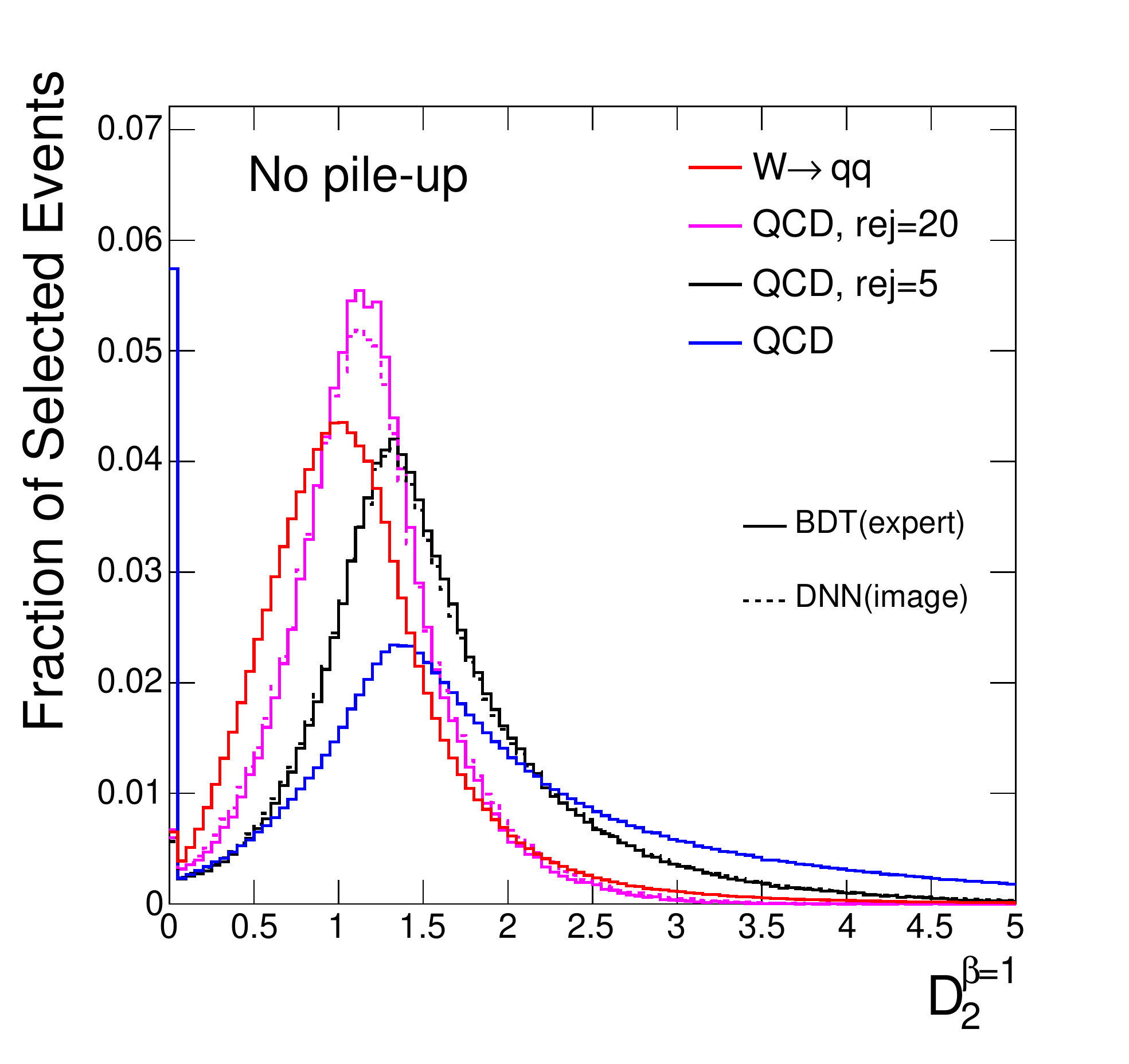}
\includegraphics[width=0.45\linewidth]{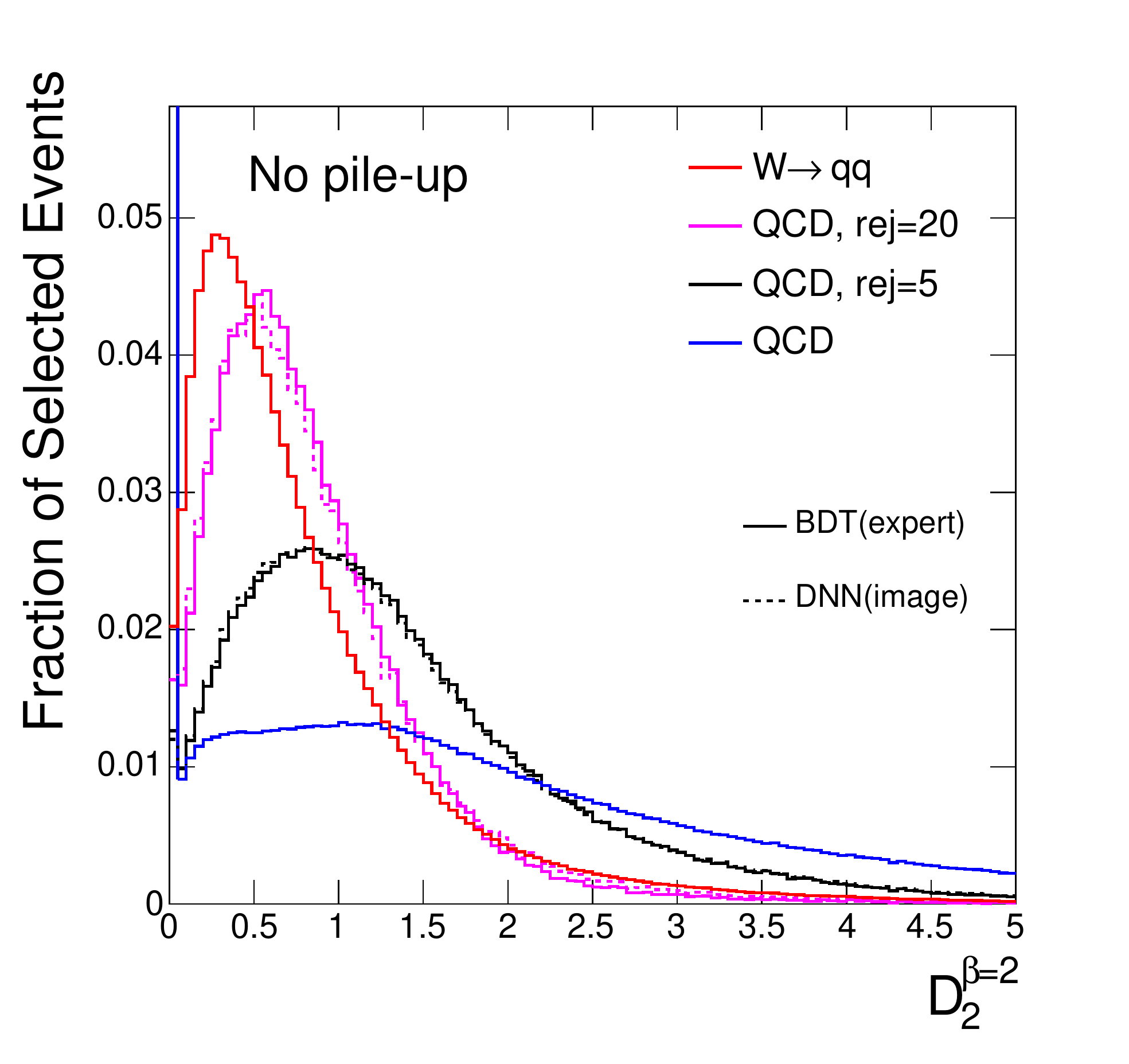}
\includegraphics[width=0.45\linewidth]{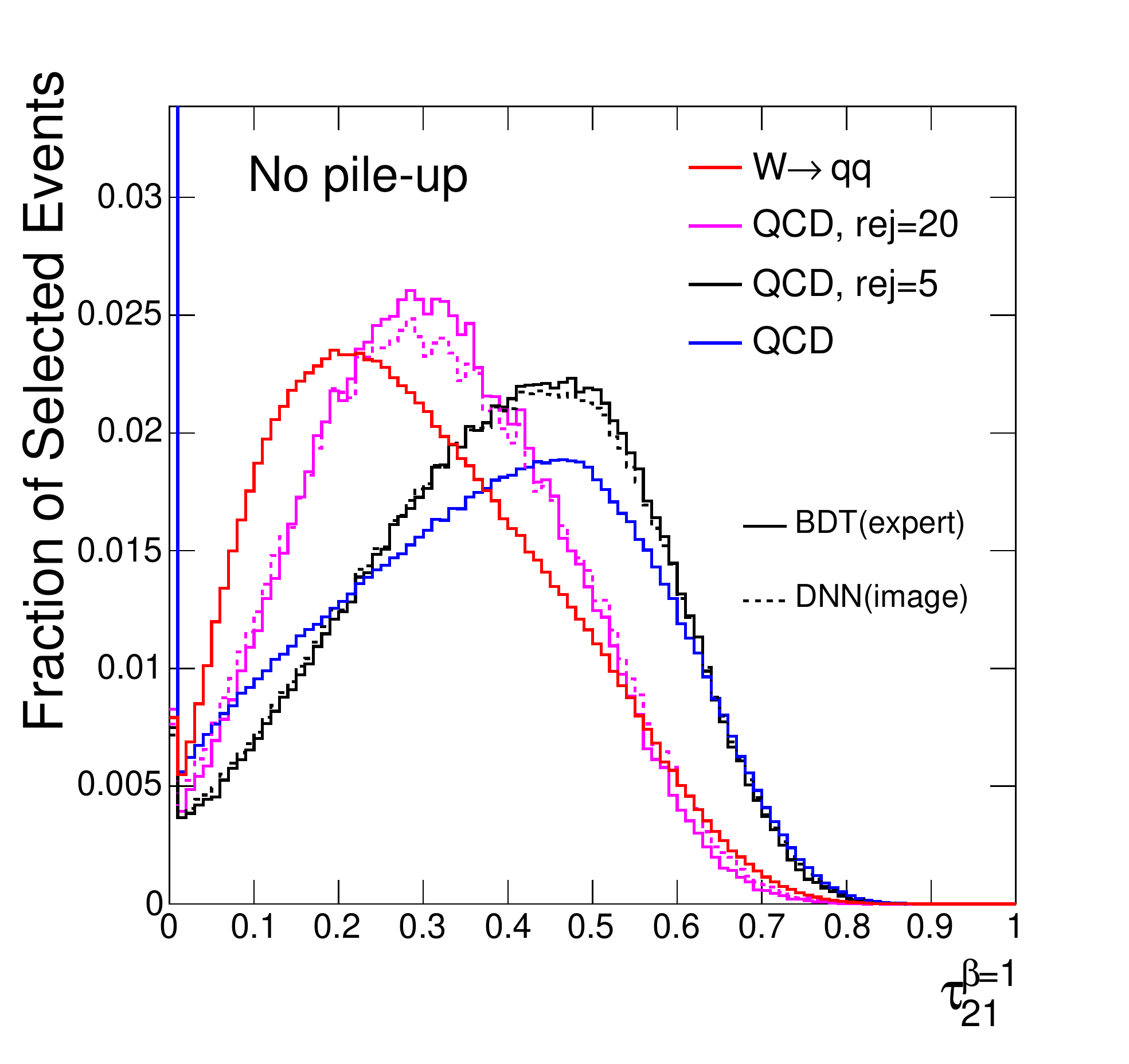}
\end{center}
\caption{Distributions in simulated samples without pileup of  high-level jet substructure variables for pure signal ($W\rightarrow qq$) and pure background (QCD) events. To explore the decision surface of the ML algorithms, also shown are background events with various levels of rejection for deep networks trained on the images and boosted decision trees trained on the expert features.  Both algorithms preferentially select jets with values near the peak signal values. Note, however, that while the BDT has been supplied with these features as an input, the DNN has learned this on its own.}
\label{fig:slice}
\end{figure}

\begin{figure}
\begin{center}
\includegraphics[width=0.45\linewidth]{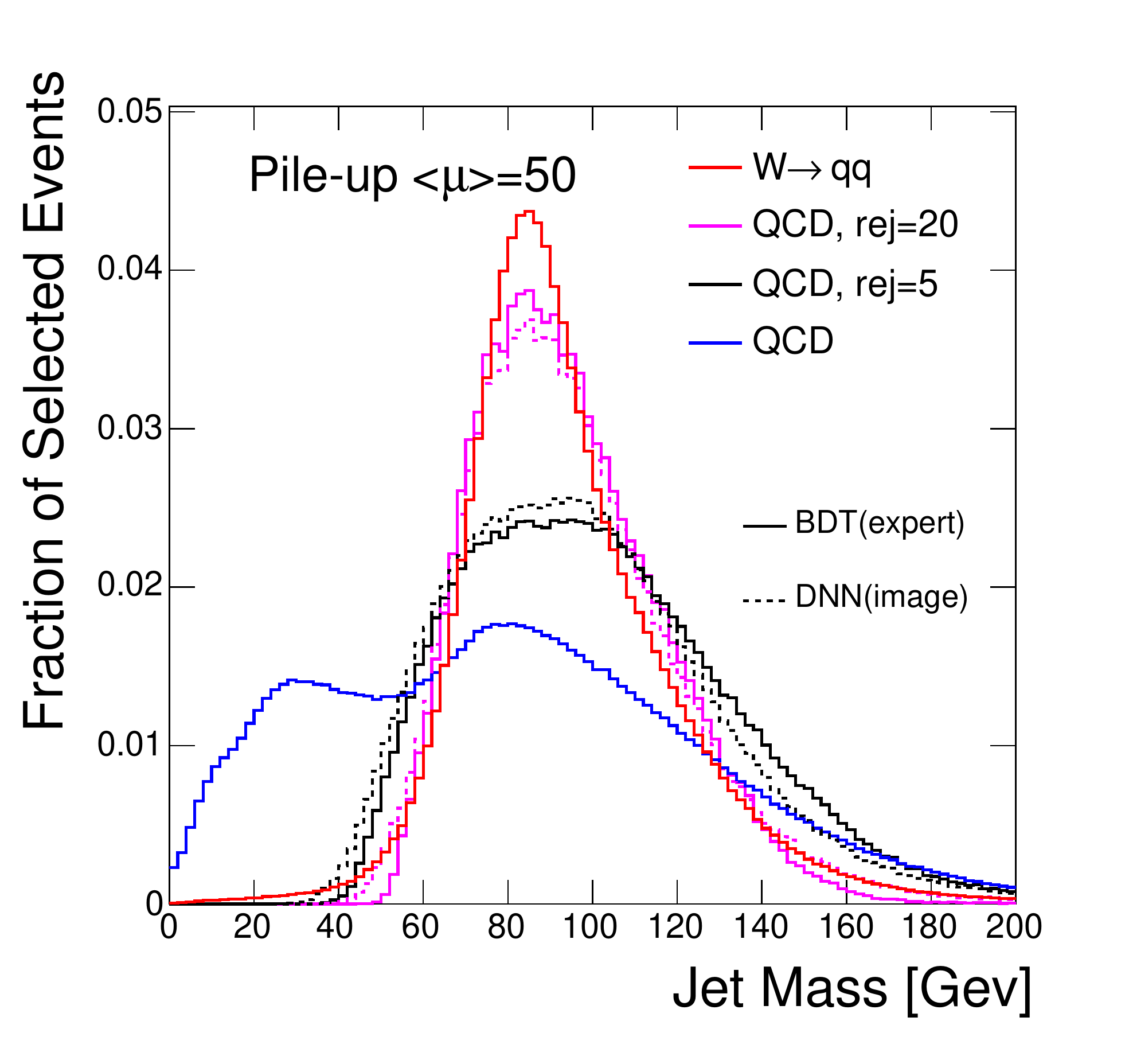}
\includegraphics[width=0.45\linewidth]{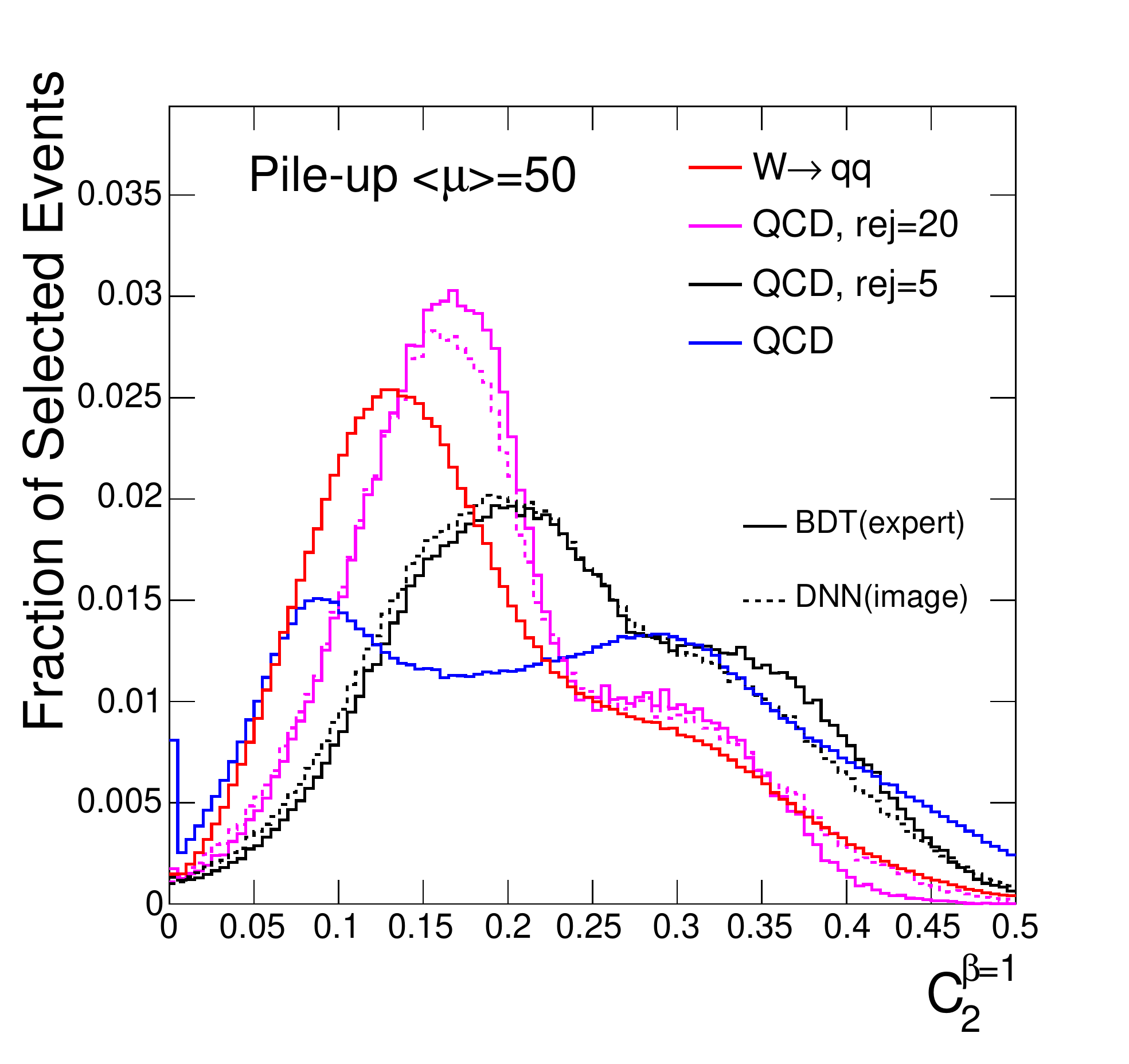}
\includegraphics[width=0.45\linewidth]{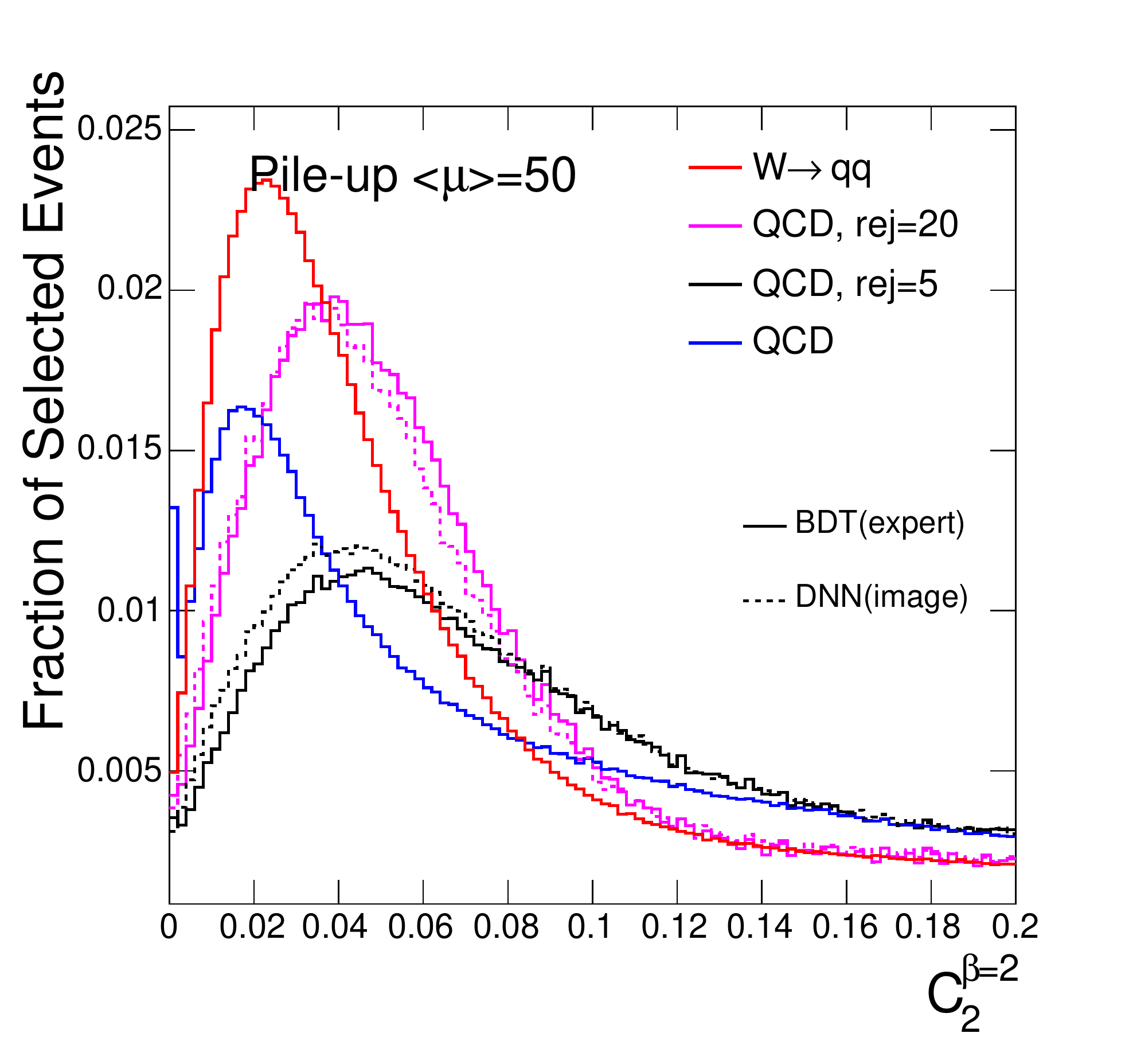}
\includegraphics[width=0.45\linewidth]{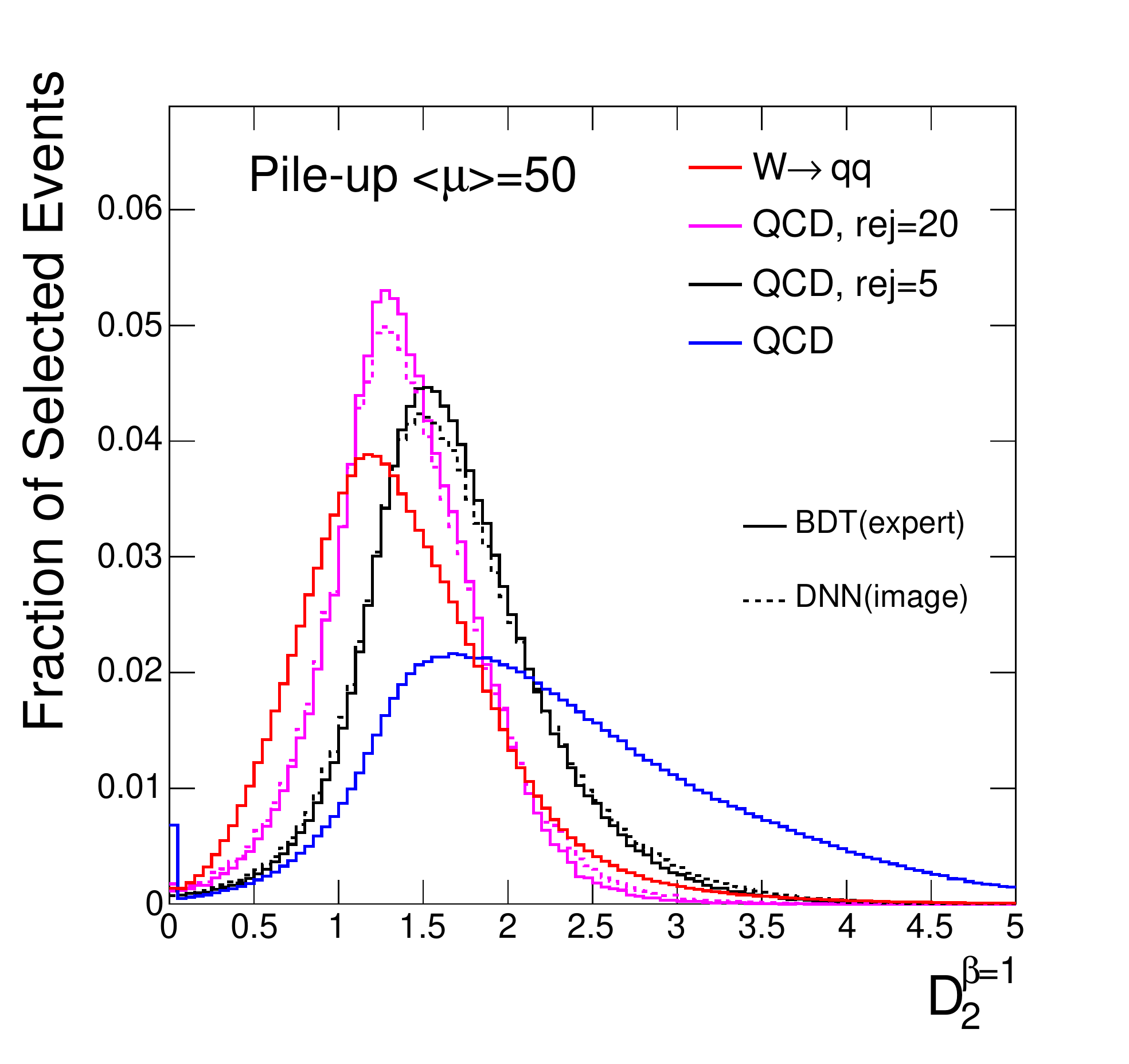}
\includegraphics[width=0.45\linewidth]{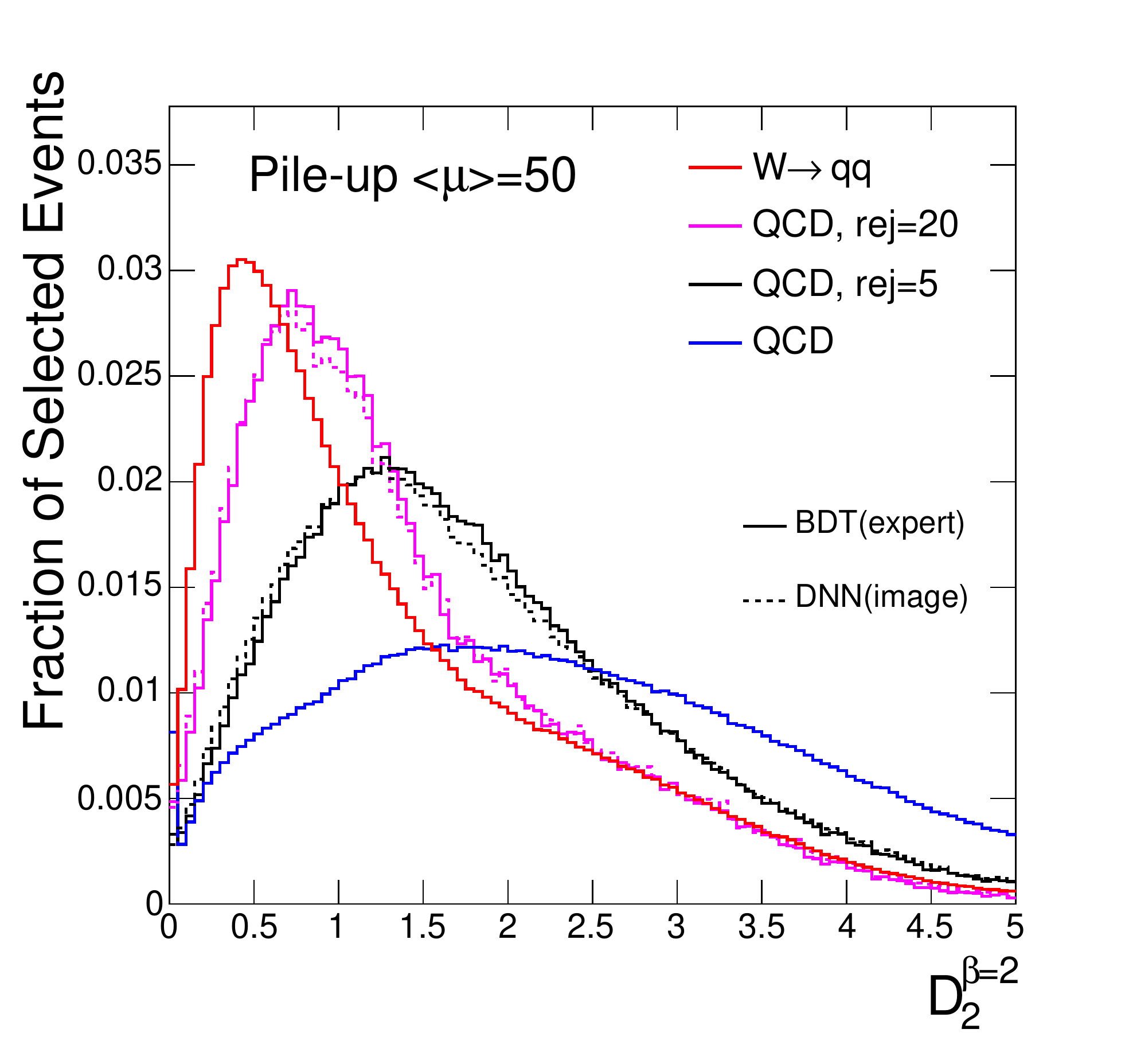}
\includegraphics[width=0.45\linewidth]{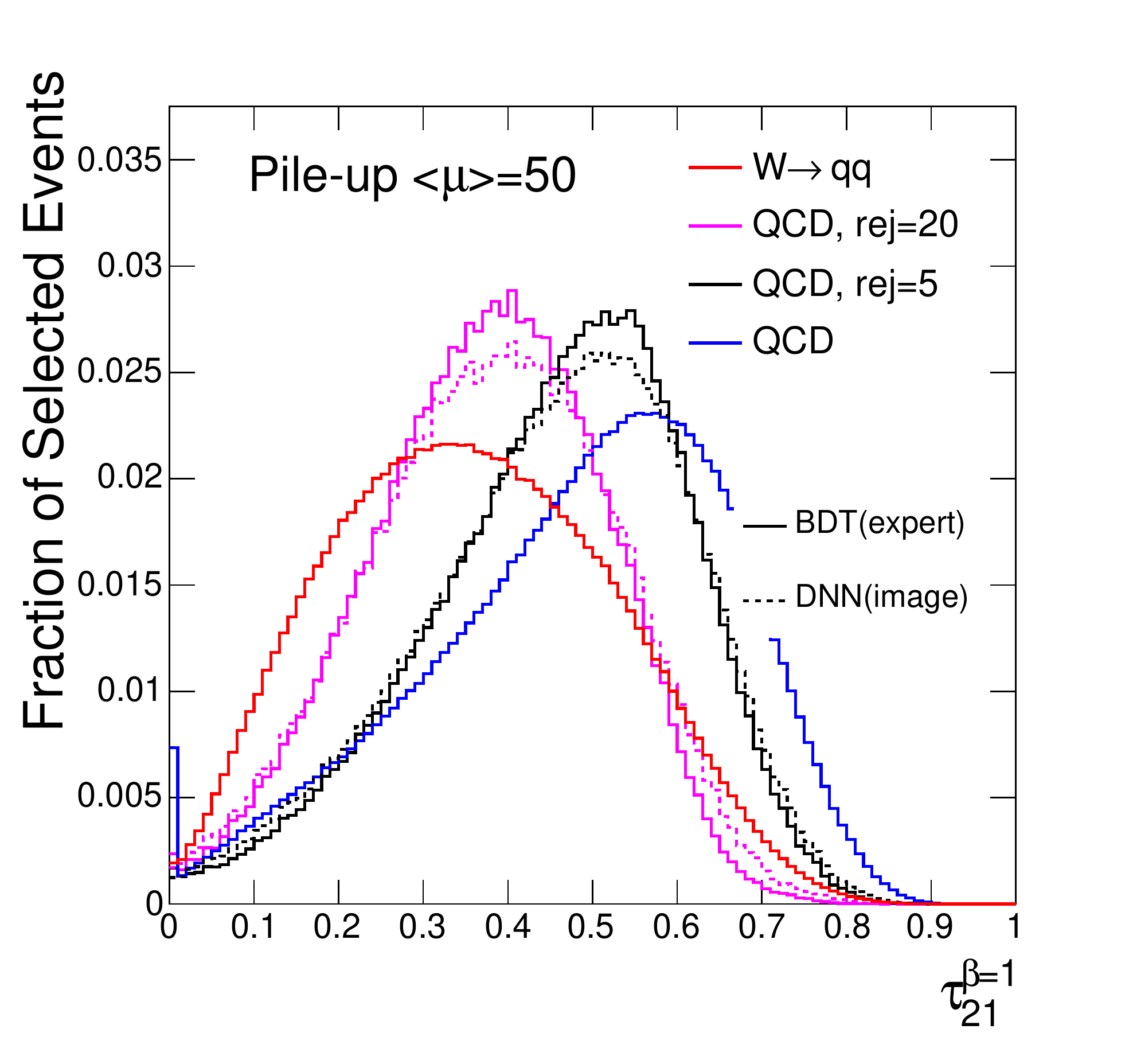}
\end{center}
\caption{Distributions in simulated samples with pileup of  high-level jet substructure variables for pure signal ($W\rightarrow qq$) and pure background (QCD) events. To explore the decision surface of the ML algorithms, also shown are background events with various levels of rejection for deep networks trained on the images and boosted decision trees trained on the expert features.  Both algorithms preferentially select jets with values near the peak signal values. Note, however, that while the BDT has been supplied with these features as an input, the DNN has learned this on its own.}
\label{fig:slice_pu}
\end{figure}

While one cannot probe the motivation of the ML algorithm, it is possible to compare distributions of events categorized as signal-like by the different algorithms in order to understand how the classification is being accomplished.  To compare distributions between different algorithms,  we study simulated events with equivalent background rejection, see Figs.~\ref{fig:slice} and~\ref{fig:slice_pu} for a comparison of the selected regions in the expert features for the two classifiers. The BDT preferentially selects events with values of the features close to the characteristic signal values and away from background-dominated values. The DNN, which has a modestly higher efficiency for the equivalent rejection, selects events near the same signal values, but in some cases can be seen to retains a slightly higher fraction of jets away from the signal-dominated region. The likely explanation is that the DNN has discovered the same signal-rich region identified by the expert features, but has in addition found avenues to optimize the performance and carve into the background-dominated region. Note that DNNs can also be trained to be independent of mass, by providing a range of mass in training, or training a network explicitly parameterized~\cite{Baldi:2016fzo,Cranmer:2015bka} in mass.

%%%%%%%%%%%%%%%%%%%%%%%%%%%%%%%%%%%%%%%%%%%%%%%%%%%%%%%%%%%%%%%%%%%%%%%%%%%%%%%%%%%%
\section{Discussion}

The signal from massive $W\rightarrow qq$ jets is typically obscured by a background from the copiously produced low-mass jets due to quarks or gluons. Highly efficient classification is critical, and even a small relative improvement in the classification accuracy can lead to a significant boost in the power of the collected data to make statistically significant discoveries. Operating the collider is very expensive, so particle physicists need tools that allow them to make the most of a fixed-size dataset. However, improving classifier performance becomes increasingly difficult as the accuracy of the classifier increases. 

Physicists have spent significant time and effort designing features for jet-tagging classification tasks. These designed features are theoretically well motivated, but as their derivation is based on a somewhat idealized description of the task (without detector or pileup effects), they cannot capture the totality of the information contained in the jet image.  We report the first studies of the application of deep learning tools to the jet substructure problem to include simulation of detector and pileup effects.

Our experiments support two conclusions. First, that machine learning methods, particularly \emph{deep} learning, can \emph{automatically} extract the knowledge necessary for classification, in principle eliminating the exclusive reliance on expert features.  The slight improvement in classification power offered by the deep network compared to the combination of expert features is likely due to the fact that the network has succeeded in discovering small optimizations of the expert features in order to account for the detector and pileup  effects present in the simulated samples.  This marks another demonstration of the power of deep networks to identify important features in high-dimensional problems.  In practice,  while deep network classification can boost jet tagging performance, expert features offer powerful insight~\cite{Larkoski:2015kga} into the validity of the simulation models used to train these networks.  We do not claim that these results make expert features obsolete.  However, it suggests that deep networks can provide similar performance on a variety of related problems where the theoretical tools are not as mature.  For example, current tools do not always include information from tracking detectors, nor do they offer performance parameterized~\cite{Baldi:2016fzo,Cranmer:2015bka} in the mass of the decaying heavy state. 

Second, we conclude that the current set of expert features when used in combination (via BDT or other shallow multi-variate approach) appear to capture nearly all of the relevant information in the high-dimensional low-level features describe by the jet image.  The power of the networks described here is limited by the accuracy of these models, and expert features may be more robust to variation among the several existing simulation models~\cite{Dolen:2016kst}.  In experimental applications, this reliance on simulation can be mitigated by using training samples from real collision data, where the labels are derived using orthogonal information.

Data in high energy physics can often be formulated as images. Thus, these results reported on the representative classification task of single $q$ or $g$ jets versus massive  jets from $W\rightarrow qq'$ are very likely to apply to a broader set of similar tasks, such as classifying jets with three constituents, as in the case of top quark decay $t\rightarrow Wb\rightarrow qq'b$, or massive jets from other particles such as  Higgs boson decays to bottom quark pairs. Note that in more realistic datasets, calorimeter information often contains depth information as well, such that the images are three-dimensional instead of two; however, this does not represent a difficult extrapolation for the machine learning algorithms.  While the fundamental classification problems are very similar from a machine learning standpoint, the literature of expert features is somewhat less mature, further underlining the potential utility of the reported deep learning methods in these areas.

Future directions of research include studies of the robustness of such networks to systematic uncertainties in the input features and to change in the hadronization and showering model used in the simulated events.

Datasets used in this paper containing millions of simulated collisions can be found in the UCI Machine Learning Repository~\cite{hepjets}. 

%%%%%%%%%%%%%%%%%%%%%%%%%%%%%%%%%%%%%%%%%%%%%%%%%%%%%%%%%%%%%%%%%%%%%%%%%%%%%%%%%%%%

\section{Acknowledgements}

We thank Jesse Thaler, James Ferrando, Sal Rappoccio, Sam Meehan, Chase Shimmin, Daniel Guest, Kyle Cranmer and Andrew Larkoski for useful comments and helpful discussion.   We thank Yuzo Kanomata for computing support. We also wish to acknowledge a hardware grant from NVIDIA and NSF grant  IIS-1321053 to PB.

%%%%%%%%%%%%
\bibliographystyle{unsrt}
\bibliography{sadowski,physics}

\clearpage
\appendix

\end{document}